A ubiquitous ~62 Myr periodic fluctuation superimposed on general trends in fossil biodiversity: II, Evolutionary dynamics associated with periodic fluctuation in marine diversity.

Adrian L. Melott and Richard K. Bambach

RRH: 62 MYR PERIODICITY: EVOLUTIONARY DYNAMICS

LRH: ADRIAN L. MELOTT and RICHARD K. BAMBACH

*Abstract.*— We use Fourier Analysis to investigate evolutionary dynamics related to periodicities in marine fossil biodiversity. Coherent periodic fluctuation in both origination and extinction of "short-lived" genera (those that survive <45 million years) is the source of an observed ~62 million year periodicity (which we confirmed in Paper I and also found to be ubiquitous in global compilations of marine diversity). We also show that the evolutionary dynamics of "long-lived" genera (those that survive >45 million years) do not participate in the periodic fluctuation in diversity and differ from those of "short-lived" genera. The difference between the evolutionary dynamics of long and short-lived genera indicates that the periodic pattern is not simply an artifact of variation in quality of the geologic record. Also, the interplay of these two previously undifferentiated systems, together with the secular increase in abundance of "long-lived" genera, is probably the source of observed but heretofore unexplained differences in evolutionary dynamics between the Paleozoic and post-Paleozoic as reported by others. Testing for cycles similar to the 62 Myr cycle in fossil biodiversity superimposed on the long-term trends of the Phanerozoic as described in Paper I, we find a significant (but weaker) signal in sedimentary rock packages, particularly carbonates, which suggests a connection. The presence of a periodic pattern in evolutionary dynamics of the more vulnerable "short-lived" component of the marine fauna demonstrates that a long-term periodic fluctuation in environmental conditions capable of affecting evolution in the marine realm characterizes our planet's history.   Coincidence in timing is more consistent with a common cause than sampling bias. A previously identified set of mass extinctions are preferentially occur during the declining phase of the 62 Myr periodicity, supporting the idea that the periodicity relates to variation in biotically important stresses. Further work should focus on finding links to physical phenomena that might reveal the causal system or systems.

*Adrian L. Melott, Department of Physics and Astronomy, University of Kansas, Lawrence, Kansas 66045  E-mail:* ***melott@ku.edu***

*Richard K. Bambach, Department of Paleobiology, National Museum of Natural History, Smithsonian Institution, PO Box 37012, MRC 121, Washington, DC 20013-7012  E-mail:* ***richard.bambach@verizon.net***

**I. Introduction**

This is the successor paper to Melott and Bambach (2011), hereafter Paper I, which dealt with a 62 Myr periodicity that exists across many independent sets of data on the Phanerozoic biodiversity of marine animals.  In Paper I, Fourier series and related techniques were used to study fluctuations superimposed upon the long-term trends in fossil biodiversity. Closely agreeing significant cycles at 62 ± 3 Myr were found in a variety of nearly independent data sets. These data sets are (a) various versions and reductions of the Sepkoski compendium of genus diversity (Sepkoski 2002), especially the reduction by Rohde and Muller (2005), hereafter R&M, (b) the Paleobiology Database genus curve (Alroy et al. 2008), hereafter PBDB, and (c) the marine family data from The Fossil Record 2 (Benton 1993, 1995), hereafter FR2.  The 62 Myr perioidicity was found not to exist at a significant level in terrestrial fauna.  The signal comes almost entirely from diversity fluctuation of "short-lived" marine genera (hereafter SLG), i.e. those that survive less than 45 Myr, the mean length of a geological period. The proportion of "long-lived" genera increases with time over the Phanerozoic, gradually reducing the signal strength in the overall sample, suggesting that some sort of "resistance" may have evolved. For details about this summary we refer you to the contents of Paper I.

To understand how origination and extinction generate the 62 Myr periodic signal in marine animal diversity it is necessary to determine their time variation and how evolutionary dynamics have interacted over time. We show here that (a) the source of the periodicity is coherent periodic fluctuation in both origination and extinction of SLG, (b) SLG evolutionary dynamics are different from those of "long-lived" genera (hereafter

LLG, those that survive >45 Myr) and (c) this substantial difference remains unchanged over time. We also argue that the interplay of these two systems is probably the source of differences observed by others in overall evolutionary dynamics between the Paleozoic and post-Paleozoic (Foote 2000; Miller and Foote 2003, 2009) and suggest that they need not have been produced by unknown secular change in global physical conditions affecting evolution or modification of evolutionary dynamics. However, the 62 Myr periodicities in diversity and evolutionary dynamics of SLG suggest long-term periodic fluctuation in conditions in the marine realm that affect the severity of extinction and likelihood of successful origination of SLG.

Fig. 1A illustrates the history of marine animal genus diversity including compilations from the Sepkoski (2002) compendium grouped by genus longevity. The major source of fluctuation in diversity of the total fauna (aside from secular increase) resides in the SLG, which we showed in Paper I are almost exclusively responsible for the ~62 Myr periodicity in diversity. Fig. 1B shows the power spectrum of detrended diversity of SLG with its dominant 62 Myr peak, which supplies 47% of the SLG variance. For the total fauna this signal virtually disappears at later times (Rohde and Muller 2005; Lieberman and Melott 2007) because SLG decrease as a fraction of the total fauna (Paper I), particularly after 150 Mya, but the signal persists undiminished for the SLG portion of the fauna.

Because the ~62 Myr periodicity is present in all comprehensive datasets for Phanerozoic marine diversity it is reasonable to expect that the data from which the periodicity was initially determined will be informative about that periodicity. Therefore, we developed our analyses in this paper from data supplied by Rohde and Muller (2005), who compiled them from the Sepkoski compendium.

## II. Methods

The periodicity signal in SLG diversity requires some form of periodicity in SLG origination or extinction, (hereafter SLGO, SLGE) or from their interaction. To avoid error from variable interval lengths when identifying trends *which integrate over time to produce diversity change* it is appropriate to use rates (number per Myr, taken to be constant over substages) even though much evolutionary turnover may be pulsed in short events within intervals. A fully detailed set of information on rates within stages, even if it could be known, would affect the spectra only at high frequencies we do not consider. Note also, that if we were exploring the intensity of extinction events, taken as impulsive events (Melott and Bambach 2010), a different metric would be appropriate.

We use Fourier Analysis, a standard method of time series analysis described in many textbooks. While nearly any series of numbers can be decomposed into the sum of a series of sinusoidal curves, some frequencies have much greater amplitudes than others and can be assigned as a significant contributor to periodicity in the series. See Melott and Bambach (2011) and references therein for further discussion.

Some choices must be made in preparing data. Detrending, the removal of long-term or secular changes from a time series and a standard procedure necessary for Fourier-based time series analysis is one. Our interest is in fluctuations around the trend, rather than the trend itself. If the trend were not removed, the results would contain artifacts which represent the trend itself, and which may obscure features of interest. Additional discussion can be found in Rohde and Muller 2005, Cornette 2007, and Paper I.  In Figure 2A we show the data on short-lived genera. The data has been truncated at 518 Ma due to small-number statistics, and at 45 Ma because after that genera cannot be assigned to SLG or LLG. Then a best-fit cubic is determined by least-squares. We find that a cubic is generally the best option for such diversity data, in that it produces the greatest reduction of residuals per free parameter of any simple functional shape. Figure 2B shows the result that is the starting-point for time series analysis.  Note that long trends, corresponding to very low-frequency information have been removed. We have also used a low level of interpolation (linear interpolation between adjacent data points to produce 1 Myr intervals). This produces a small effect at the highest frequencies we consider, and a negligible one at others, as shown in Paper I.  The mean interval length is 3.2 Myr, with the longest 9.45 Myr. In order to take advantage of greater generality, our results are presented here based on standard Fourier Analysis, using AutoSignal 1.7 and software written by Melott. They have been checked as before (Cornette 2007; Lieberman and Melott 2007; Melott and Bambach 2011) by using Lomb-Scargle transform (Scargle 1982), which does not require interpolation.

Another choice concerns computing the power spectra of the origination (O) and extinction of (E) of the short-lived fauna, SLGO and SLGE.  These were computed using the rates, defined as the number of O or E in an interval, divided by the length of the interval in Myr. The choice of rates as numbers of events per Myr rather than numbers of events per interval has a measurable effect. We recognize that there is considerable evidence that these events are pulsed (Foote 2005), so the rate is far from constant within intervals. The power spectra of rates of SLGE and SLGO both show peaks at ~62 Myr, which are, however, not significant against the background (Fig. 3A, 3B).

Their lack of formal significance against the background is caused by the fluctuations in the right-hand third of the area covered by Figures 3A and 3B. These are the short-period (roughly <15 Myr), high-frequency oscillations which are emphasized by taking a derivative (rate). Although this is a mathematically based statement, it may be understood intuitively by noting that the same sized change will have a larger rate if it is the result of a higher frequency (shorter period) oscillation. Extinction and origination change biodiversity, and when we look at this plot we are seeing fluctuation in the rate of change—twice removed from the data.  The relatively high amplitude changes from interval to interval dominate the fit to the spectrum and raise the overall significance level demanded.  It may be instructive to compare the overall slope of 3A and 3B to that of the spectrum in Figure 1B or to the spectra in Paper I.

The use of rates is conservative with respect to our conclusions. We have examined the distribution of lengths of intervals as a function of the number of O or E events in those intervals. We find that these quantities tend to increase, but less than linearly.  If the increase were linear, the fluctuations in rates would be strongly suppressed.  If there were no trend, then rates and absolute numbers of events would give equivalent results.  Given the sublinear trend, if we use absolute numbers rather than the rates we use, it magnifies the apparent fluctuations. The ~62 Myr periodicity for number per interval is more pronounced in both SLGO and SLGE than it is in their rates (per Myr), and SGLO is highly significant in that evaluation. By comparing the distribution of the data against the periodicity of interest it is apparent that other variance in the data is quite large and this reduces the dominance of the ~62 Myr periodicity in the power spectra of SLGE and SLGO rates, but visual comparison of SLGE and SLGO to their ~62 Myr spectral peaks does reveal a correlation. As discussed elsewhere it is important that the area under our curves be equal to the number of O or E, and that they be spaced in time, not stage, or we cannot correctly use them in time series analysis to determine their contribution to biodiversity and its changes over time.  While the events may be pulsed, the assignment of a pulsed value to events of finite separation would incorrectly assign a greater contribution to the spectrum by those occurring within longer stages. This makes spectral peaks more significant, but we do not consider this a correct procedure given our goals. For similar reasons, in this study we do not focus strongly on proportions of genera involved in extinction or origination (intensity of change relative to extant numbers of genera), but it gives interesting results on the severity of extinction (Melott and Bambach 2011). Also, both O and E rates are assigned to the mid-point of stages, rather than to their boundaries as in some other studies. We have checked and determined that this choice (compared with assigning all O to the beginnings of stages and all E to the ends) does not significantly affect our spectra, but it results in a displacement of our cross-correlations shown later by about 3 Myr, the average stage length, exactly as discussed in Kirchner and Weil (2000). In terms of spectra, it would noticeably affect only high frequencies which we do not consider.  Thus this choice makes a shift within the typical measurement errors for periods (± 3 Myr) in existing studies of the 62 Myr cycle; at no point are our conclusions dependent upon this choice.

So, if the spectral peaks in SLO and SLE separately are not strong enough to construct the major peak in biodiversity seen in Figure 1B, how does it arise?  The answer is that for the short-lived genera, and not for the long-lived genera, and only around the 62 Myr wavelength, the timing of the peaks of O and E is such that they combine to make strong oscillations in biodiversity.  This will be seen in the coming discussion, and particularly in Figures 4A, 4B, and the related discussion. Although the amplitudes we have seen are not extremely high, the phase relationship of these peaks makes significant cumulative change in O-E, or biodiversity.

### III. Evolutionary dynamics of SLG

A. *How the interaction of origination and extinction generates changes in biodiversity in SLG.*— The power spectra of SLGE and SLGO rates both show peaks at ~62 Myr, which are, however, not significant against the background (Fig. 3A, 3B). The periodic changes in SLG diversity depend upon the interaction of SLGE with SLGO, best described by the cross-spectrum of SLGE and SLGO, a complex number generalization of the power spectrum. In the past we have plotted only the real part of this quantity, which is a measure of the extent to which waves are in phase (i. e., reach their peak at the same time). Now, since O/E may lag/lead one another, we plot the amplitude, labeling the phase lag of our peaks of interest in Myr. The amplitude verifies that the interaction of these fluctuations produces by far the strongest signal at a period of 62 Myr (Fig. 4A).  The complex phase angle gives different information, and shows that SLGE lags SLGO in this periodicity by about 14 Myr (Fig. 4B).  Note that at high frequencies the phase angle fluctuates wildly, characteristic of noise, and that at lower frequencies the lag is reversed (i. e., remains positive). Only near the strong periodicity of interest is the lag coherent and negative.

Fig. 4C shows the relationship between SLG diversity, SLGO, and SLGE.  Note that we have truncated the diversity data at 44 Ma, after which genera cannot be separated into SLG or LLG. The figure is for detrended data in which negative values are for diversities and rates less than the trend line. The coherent interaction of the periodicity in SLGO and SLGE, not fluctuations in just one or the other, produces the 62 Myr periodicity in SLG diversity. The best-fit peaks of SLGO fall 8 Myr before SLG diversity peaks, and SLGE peaks fall 6 Myr after. In analyzing data for the total fauna Rohde and Muller (2005) observed that both origination and extinction displayed a 62 Myr periodicity, but they did not determine the nature of the interaction between them that drives the periodicity in diversity.

B. *Intraction produces consistent diversity fluctuation, but neither O or E alone always shift consistently.*— Nine full or partial cycles of the ~62 Myr periodicity in diversity of SLG occur between 540 and 44 Mya which also permits us to evaluate the way origination and extinction of SLG interacted to produce each phase of increase and decrease of diversity in each cycle. We emphasize that the strong significant diversity periodicity is a result of the *interaction* of SLGO rate and SLGE rate — a consistent driver is not present in either alone. This can be understood by comparing the behavior of SLGO and SLGE rates during times of increase in diversity with those during times of decrease in diversity. During times of increase in diversity SLGO rate must exceed SLGE rate and during times of decreasing diversity SLGE rate must exceed SLGO rate, on average. Shifts in one or both rates must occur in the transitions between phases of increasing and decreasing diversity to cause fluctuation in diversity.

In three of the eight complete diversity cycles tracked through the Phanerozoic, decreases in diversity were produced by a distinct decrease in SLGO rate coupled with a distinct increase in SLGE rate compared to the preceding interval of increasing diversity, as would be expected if regular fluctuation in both origination and extinction rate were always involved in driving the periodicity. But in four cycles diversity decrease was produced in two instances by decrease in SLGO rate with no marked change in SLGE rate and in the other two instances by marked increase in SLGE rate with not much decrease in SLGO rate compared with the preceding interval of increasing diversity. In the remaining case a large increase in SLGE overwhelmed a marked increase in SLGO. Although mass extinctions primarily participate in phases of diversity decrease (see Section V below), their effect is not sufficient to drive the periodic decreases in diversity. Diversity loss from mass extinction events was needed to produce the low diversity trough of the periodicity in only three of the eight cycles.

Similarly, in only two of the eight diversity cycles was the phase of increase in diversity produced by coupling marked increase in SLGO rate and marked decrease in SLGE rate compared to the preceding phase of diversity decrease. In three cycles diversity increased because SLGO rate increased, but without a marked decrease in SLGE rate (in two cases SLGE actually increased) and in the three other cycles diversity increase resulted from a marked drop in SLGE rate without an increase in average SLGO rate when compared to the preceding phase of diversity loss. In those three instances SLGO either remained about the same or decreased somewhat. Mass extinctions did not control any of the shifts from diversity loss to diversity increase.

## IV. Contrasting evolutionary dynamics between short-lived and long-lived genera

The timing pattern for evolutionary dynamics is markedly different between SLG (Figs. 3, 4) and LLG (Fig. 5). Although a ~62 Myr peak exists in both E-O cross spectra, it is dominant for SLG (Fig. 4A) but does not rise above the other peaks for LLG (Fig. 5D). When SLGE and LLGE are cross correlated (Fig. 6A) the peak is at zero lag, indicating that both respond at the same time to the same extinction events, but far less for LLG than for SLG, as can be observed in from the stability in diversity of LLG seen in Fig. 1A. Origination, however, shows *no* similarity between SLG and LLG. In fact they tend to avoid one another (Fig 6B). Despite the fact that both SLG and LLG share the timing of response to extinction events, a strong periodic pattern like the interaction of origination and extinction in SLG is not present in LLG. For SLG there is a correlation of origination rates with extinction rates 16 Myr later (Fig 6C) and no significant correlation of SLGE following SLGO at any lag (Fig 6C right half). This behavior results in large fluctuations in SLG biodiversity, but tends to cancel out any long-term growth. LLG show a strong correlation of LLGE rates with LLGO rates 10 Myr later as well as an anticorrelation (negative correlation) between origination rates and extinction rates 8 Myr later (Fig. 6D). Both of these correlations tend to promote growth in LLG biodiversity, and suggest that it might be stimulated by mass extinctions, consistent with the sense of Krug et al. (2009). We also have partitioned the data, and examined Paleozoic and post-Paleozoic data separately. The qualitative characteristics of orignation and extinction do not change for either SLG or LLG between the Paleozoic and the post-Paleozoic. SLG and LLG have consistently different patterns of evolutionary dynamics throughout the Phanerozoic. These results may appear to be inconsistent with the report by Kirchner and Weill (2000) of enhanced origination following extinction for all genera, but they used proportions rather than numerical rates, which accounts for the apparent differences (as we confirm and discuss later).

In Fig. 5 we explore more fully the independent behavior of the LLG. The spectrum of their diversity (Figure 5A) shows a peak at 58 Myr which almost reaches the p=0.05 confidence level. The power spectra of LLGO and LLGE (in 5B and 5C) feature peaks even less dominant than analogous ones in SLG. There is some excess joint probability, as seen by the cross-spectral peak in Fig. 5D. The 60 Myr peak is the largest one, but still comparable to the others. Clearly, the long-lived genera do not participate in the main cycle with much intensity.

We now present some additional details to supplement our main points. The first is shown in Figure 7, which shows the best-fit ~62 Myr component to SLG diversity, extinction rate, and origination rate, in Figs. 7A, 7B, and 7C respectively. The size of the sinusoidal component is scaled to show how much of the variance of the main data is contributed by it. In all three of these cases, there is a tendency for the data to go up or down when the sine wave goes up or down. This is not the only variation, because the data is a sum of many trends. In parts B and C (rates) the fit is less precise because there is more high frequency variation, a result of taking differences, and as reflected in the significance assignments shown in Fig. 3A and 3B.

One of our more interesting results is the differing behavior of LLG. As an introduction to this, we show in Fig. 7D the detrended origination rate for LLG—and superimposed on it the sine wave fit for the SLG. There does not appear to be much correspondence, which introduces our results on the LLG.

It is easier to display commonality by examining cross-correlations in this case. Fig. 6A and 6B display cross-correlations between extinction rates (a) and origination rates (b) of SLG against LLG. It is important to recall that correlations include the combined effects of all parts of the spectra. In Fig. 6A (extinction) there is a strong peak at 0 lag which indicates that SLG and LLG see the same extinction events at the same times. We have confirmed this is also true for the 62 Myr cycle by noting a nearly identical phase angle in the two E spectra at this point.  In 6B (origination) we note lower amplitude features. The depression around -20 possibly indicates a mild depression of LLGO about 20 Myr before SLGO peaks.   Spectral analysis confirms that for the 62 Myr cycle, there is a tendency to avoid SLG originations: the phase angle for LLGO around 62 Myr is out of phase with SLGO.  To summarize, LLG are much less affected by the 62 Myr diversity cycle; their small participation is driven by their partial participation in mass extinction events.  This can be seen in Fig. 1A, where their diversity shows small dips at most of the mass extinction events, with the only large ones at the Permo-Triassic and end-Cretaceous events.  Insofar as their origination is keyed to this cycle, it is out of phase with the origination of SLG. Figure 8 shows the best-fit to the 62 Myr cycle in LLGO against the data. In comparison with Figure 7D which contains the SLGO sine-wave fit it can be seen that the sine-wave fits are very nearly inverted.  While SLG tend to originate just before peaks of their diversity, LLG originate most often at or just after their minima. The reader may have noticed that the sine-wave amplitude in Fig 8 grows with time, which is possible because it represents a narrow band of frequencies, not a single frequency. The amount of LLG origination in the wake of mass extinctions also appears to grow with time. This is in strong accord with the result that genera which originate just after times of mass extinction during the post-Paleozoic are more long-lived (Miller and Foote 2003).

## V. Relation of the 62 Myr Cycle to extinction events

*A. Distribution of extinction events against the 62 Myr cycle.*— Our primary interest here is the time dependence of fossil biodiversity, not the timing of extinctions. However, extinctions obviously affect biodiversity and we may test for the validity of certain expectations derived from the results described earlier.

We tested a list of 62 extinction events, most of which are identified in Barnes et al. (1996) (many of which are also mentioned in Hallam and Wignall (1997)) and with one added from Hallam and Wignall (1997) and one from Sepkoski (1996). These include the 19 mass extinctions defined by Bambach (2006) as well as 43 additional recognized extinction events that do not meet all the criteria (peaks in both magnitude and rate of extinction in all data compilations examined) for designation as mass extinctions as established in Bambach (2006). As these lists were constructed and published in a fully independent analysis of the fossil record without taking account of any claims to periodicity, we check them here for consistency against the periodicity fits.

If the periodic cycles are "real", in the sense that they capture something going on in biodiversity, then there should be some correlation with extinctions. Our biodiversity periodicity could have some relationship to extinction events in the following way: such events should happen preferentially on the "downturn" side of the cycles, when biodiversity is dropping.  Therefore our null hypothesis is:  "There is no correlation between the timing of this list of extinctions and a negative slope to a given significant component of the biodiversity cycle."

We perform this test as follows: we note that there is a 50% probability that a random, uncorrelated event will occur when a given sinusoid is on the way up rather than on the way down.  We use the sine component fits to the R&M data, the PBDB data as analyzed in Melott (2008), and the FR2 marine biodiversity fit performed in Paper I.  These fits are of the form

$$C = A \sin (f\, t + \varphi) \tag{1}$$

where C is the contribution to biodiversity from the given component, A is the amplitude of the wave, f is the frequency in "per Myr", t is the time in Ma, and φ is a phase angle that determines the position of the cycle.  We are only interested in the sign of the slope of this function; its slope is the derivative, cosine. In Figure 9A, we plot the three different fits for the 62 Myr cycle: the value from the R&M data, which is f=0.01609, φ=5.193, the PBDB values, f=0.01589, φ=5.416, and the FR2 marine genera data f=0.01642, φ=4.512. Note that these fits are written with the sense of time as "into the past", Ma. The plots, however, show Ma as negative numbers to indicate time before the present. The shaded areas in Fig. 9A correspond to times when the 62 Myr component of the R&M fit to detrended biodiversity is in decline toward the present. The slight differences in the three curves are consistent with the cross-spectral analyses discussed in Paper I.  Note that all three curves coincide near the end-Permian event, which may be an "anchor" for the fits by virtue of the large amplitude of the biodiversity change.  Interestingly, the total set of 62 extinction events is split evenly between the increasing and decreasing diversity phases of the 62 Myr periodicity determined from the R&M data set, and is not far from this for the other two (Fig. 9A).  Thus this list of extinction events seems to have no particular relation to the 62 Myr cycle.

   *B. 19 mass extinctions identified by Bambach.*— A contrasting result is provided by the timing of 19 mass extinction events identified by Bambach (2006).  These events, which are a subset of the total of 62 extinction events listed in Barnes et al. (1996) were selected on the basis of two criteria: (1) the intensity of extinction (the rate in the given interval) and (2) the magnitude (total amount of extinction in the given interval). A typo is corrected for the end date of the late Guadalupian extinction, and the complex Ashgillian (Late Ordovician) extinction is entered as two separate events, because it is clearly resolved as such in the fossil record (e.g. Bambach 2006, and references therein). Substituting the Bambach (2006) data for the values of t, we examine the 19 mass extinction events. These are shown as the black filled circles in Figure 9B. 15 are on the declining half of the 62 Myr cycle of the R&M fit, 14 on the declining half of the PBDB fit, and 13 on the declining half of the FR2 fit. These are all more than half (9.5) which is the expectation value if they were unrelated to the 62 Myr cycle. The proper way to (individually) characterize these results is by the probability of having no more than 4 which are on the increasing side (R&M), which corresponds to p=0.0095, no more than 5 on the increasing side (PBDB), p=0.0317, or no more than 6 on the increasing side (FR2), p=0.0835.  Thus, we can reject the null hypothesis with confidence only 92% on the basis of the FR2 data, but 97% on the basis of PBDB data, and 99% with R&M data. It is not possible to generate a meaningful joint probability without detailed information of the degree of dependence between the data sets. If we approximate them as completely independent, the p-value for finding all three asymmetries in mass extinction is 0.000025.  Since FR2 is somewhat dependent on Sepkoski (2002), which used some of the same data on genus end points, and which was used as a source for R&M, if we throw out FR2 and only consider R&M and PBDB as independent samples, we have p=0.0003, or 99.97% confidence in the rejection of the null hypothesis that it has nothing to do with any of the fits.  Therefore, these special "mass extinction events" are clearly different.  We can reject the null hypothesis that they are unrelated to the three similar ~62 Myr cycles with very high confidence.

*C. Extinctions that did not qualify as "mass extinctions".*--We can also separately examine the "additional extinction events", i.e. those on the list of 62 referred to in the previous sections that do not meet the criteria for inclusion in the list of 19 identified by Bambach (2006) as mass extinctions. However, just as the subset of 19 mass extinction events defined in Bambach (2006) are preferentially clustered in the decreasing diversity phases, 27 of the 43 "additional extinction events" lie on the rising side of the 62 Myr cycle and just 16 fall in the decreasing diversity phase (Fig. 9C). (Note that there are two separate biases, the one found in the previous section for the mass extinction events, an additional and opposite one for these events, and no bias at all for the set of 62 as a whole.) Using the normal approximation to the binomial distribution, the bias for the "additional extinction events" is better than 1.68 σ, representing a deficit from what is expected at the p=0.05 level. It is possible that some of these may represent "failed mass extinctions" in some sense.  This result may relate to phenomena such as those hypothesized by Arens and West (2008), that pulse stresses alone may seldom cause extinction to reach "mass extinction" levels (sensu Bambach 2006), but when coupled to elevated background stress, which may typify the decreasing diversity phases of the 62 Myr periodicity, large scale extinction events are more frequently triggered.

We also used this method to look at the 157 and 46 Myr signals found in the PBDB.  They showed no significant deviation from what would be expected to occur frequently in a 50-50 choice.  Thus, this test, like our others is consistent with the idea the 62 Myr cycle is capturing a significant aspect of biodiversity history, but does not similarly support the other possible signals.

Elsewhere we have noted that a 27 Myr periodicity in extinction intensity shared between the Sepkoski and PBDB data sets is detectable and that an unusual number of mass extinction events occur with that temporal spacing (Melott and Bambach 2010). Since mass extinctions are not the sole control of the extinction side of the 62 Myr diversity fluctuation we have not explored that shorter term cycle in this paper, but it is worth noting (a) that the improvements in the geological time scale since Raup and Sepkoski (1984) suggested a 26 Myr periodicity in peaks of extinction over the last 250 Myr now reveal that it exists through the entire Phanerozoic and (b) that the timing precision over the last half-billion years is such that the hypothesis of Nemesis (a dark companion to the sun with an eccentric orbit), which would have notable variation in its orbital behavior, is not observationally supported.

**VI. Variation in the amount of the sedimentary record**

*A. Concern.*-- The idea that observed diversity may be related to the variation in

the availability of sedimentary rocks has been raised many times (Peters and Foote

2001; Smith and McGowan 2005, 2007). Here we analyze the best available data set on variation in the sedimentary record (Peters 2005, 2006a, 2006b, 2008a, 2008b), and also look at efforts to evaluate sea-level through time (Lugowski and Ogg 2010, Wilgus et al 1988, Haq et al 1987). Although a relationship emerges, we argue in the following section, that it relates more to "common cause" (Peters 2006b, 2008a) than to record bias.

*B. Power spectra of sedimentary packages*--In this subsection, we present an

analysis of the number (area and temporal-discreteness weighted) of sedimentary packages documented from North America as presented by Peters (2008a).  We study carbonate packages, which are predominantly marine in origin, then siliciclastic packages, which, although sediments explicitly identified as non-marine were excluded,

probably still include some terrestrial stream, floodplain, and lake deposits along with

the marine deltas, shoreline and shelf siliciclastic deposits (Peters 2006a).  We use the

data as provided by Peters, detrended and analyzed for periodicity. We shall refer to CP

for carbonate packages and SP for siliciclastic packages.

In Figure 10A, we show the power spectrum of CP as a function of time.  A

strong peak shows up at 58 ± 4 Myr, with a confidence level p < 0.001, and it accounts

for 22% of the total variance in detrended CP.  In Paper I we found the peak in

biodiversity at 62 ± 3 Myr accounted for 47% of the variance in the Sepkoski (2002)

data. In general, we will always comment on whatever spectral feature we find close to this frequency, and if any other large ones exist, we will mention them. However, our focus is on finding relationships with the biodiversity fluctuation at about 62 Myr.

Figure 10B shows the power spectrum of SP. There is a peak at 58 ± 3 Myr, but it is of lower amplitude than with the CP, and just reaches p = 0.01.  The 58 Myr spectral peaks in CP and SP have phase angles that are very nearly equal, suggesting a common oscillation. The frequencies for CP and SP are close but not identical to the frequency in the biodiversity studies.  The frequencies also are about 1 standard deviation apart in their joint distribution. For this reason, it is worthwhile to investigate their differences and joint distribution.

*C. Cross-spectra between sedimentary packages and fossil biodiversity*-- Given the existence of spectral peaks in the record of CP and SP that resemble the prominent 62 Myr peak in fossil biodiversity justifies doing a cross-spectral analysis to search for common frequencies in phase.  In Figure 4, we looked at the amplitude of the cross-spectrum, which tells us how much there are common frequencies present in two different time series, without regard to whether they were in phase. When we plot only the real part, as we do now, the value of the function at a given frequency shows how much they jointly oscillate (at that frequency), and the extent to which they are out of phase—opposed—is shown as a negative value. The cross spectra are constructed based on the detrended CP and SP series from Peters (2008).  All series are divided by their standard deviation to put everything on an equal footing.

Figure 10C shows the real part of the cross spectrum between fossil biodiversity (using the R&M data) and CP. The overall "noise level", i.e. low-level oscillations about zero, is somewhat higher than seen before. A 60 Myr peak is prominent, and the phase angles indicate a very small 0.6 Myr difference in average timing.  A secondary peak shows up at 172 Myr with a moderate (9% of period) 15 Myr mismatch in timing.  This is strong evidence that there is considerable commonality between the number of CP and the fossil biodiversity record. The 60 Myr peak is the largest peak here, and lies within one joint standard deviation of the frequency of peaks found in biodiversity.  Note that this is a measure of its common strength between the two time series. If this peak were removed, the net area under the curve in Figure 10C would be close to zero (note the offset zero-point of the y axis.) This means that, aside from the secular trend, most of the similarity between biodiversity fluctuations and CP is captured by this oscillation.

The results for SP are somewhat different. Figure 10D shows the cross-spectrum with SP and the R&M biodiversity data.  The same spectral peaks appear as with CP, but are noticeably weaker.  The timing agreement of the larger peak is still good, but not as precise as seen between the various biodiversity time series, nor between them and the CP.  We note now that a negative peak is seen at 135 Myr, stronger in Figure 10D. This is interesting but somewhat off-topic, since our focus is the 62 Myr periodicity.

The stronger congruence of the CP with biodiversity is reasonable because CP are predominantly marine in origin and the biodiversity signal is rooted in the marine fauna, whereas Peters' SP data probably includes some terrestrial deposits, since all sediments in the source material were not identified fully as to depositional environment (Peters 2006a). This difference may be reinforced because the total fossil biodiversity fluctuation power is also stronger in the Paleozoic and early to mid Mesozoic (Lieberman and Melott 2007), when the CP are also more numerous (Peters 2008a). Because the SP contain some proportion of terrestrial deposition, as well as estuarine and deltaic deposits, which have limited, low diversity faunas, along with fully marine deposits and the terrestrial fauna apparently lacks the biodiversity signal of interest, it is reasonable that the signal is in weakened form in SP.

We wish to make it clear to the reader that Peters has reported extensively on the correspondence of diversity fluctuation and the fluctuation in number of sedimentary packages (Peters 2005, 2006b, 2008a, 2008b). What we are reporting new is on the presence of a temporal periodicity in Peters' data that relates to the 62 Myr periodicity in biodiversity, a phenomenon that Peters did not address.

D. *Periodicity in sea levels?*— Given a signal in sedimentary rock packages, it is natural to consider the possibility of a connection with sea levels.  In what follows, we use the current standard data on short-term (third order) sea-level changes for the Phanerozoic based on the data of Hardenbol et al. (1998) and Haq and Schutter (2008). The data werekindly supplied by J. Ogg and are available as part of the data used in TimeScale Creator (Lugowski and Ogg 2010) available at

https://engineering.purdue.edu/Stratigraphy/tscreator/index/index.php, hereafter referred to as TSC.

Significance levels mostly fall above the visible frame and are barely visible in the power spectrum of the TSC data in Figure 11A,.  In short, there is no evidence for any departure from the expected level excursions of Gaussian fluctuations. However, the broad peak at 58-67 Myr does have rough phase agreement with the biodiversity, as we shall see.

Figure 11B explores the cross spectrum of the TSC sea levels with the biodiversity data compiled by Rhode and Muller (2005) from the Sepkoski compendium, the data from which the 62 Myr periodicity was originally extracted.  Note that the y axes are shifted from other cross-spectral plots. There is a weak peak at 60 Myr, with excellent phase agreement.  It is reasonable within the errors to regard this as consistent with the 62 Myr periodicity in biodiversity.  However, it is important to note that *we would never have selected it for examination as important without the prior information, i.e. that we were looking for a peak in this region due to the biodiversity results.*

The cross-spectrum of the TSC sea levels with the sum of the two main types of sedimentary rock packages is shown in Figure 11C. For this cross-spectrum we used CP and SP combined because CP decreases markedly in abundance over time so the combination better represents the sediments that might have responded to sea level. As one would expect, there are indications of strong joint fluctuation in the same direction over long time periods. There is a peak at 60 Myr with phase indistinguishable from that of the biodiversity fluctuations.

We also did cross-checks with older data sets: Haq et al. (1987), for which the data was kindly provided by K. Miller, and with the Exxon sea levels (Wilgus et al 1988). It is important to note that the former data exists for a significantly shorter time period, only back to 245 Ma, and that neither has been corrected for the Gradstein (2004) timescale. Nonetheless, we saw results on the periodicity in these older data broadly consistent with that in the TSC data, but of slightly lower amplitude.

The picture which emerges from sea level is consistent with there being a low-amplitude fluctuation consistent within one standard deviation with the 62 Myr periodicity that was found in fossil biodiversity and in sedimentary rock packages. This is not something that would ever have caught our attention on its own, due to small amplitude. However, our examination of sea levels could have found inconsistency: no peaks near the period of interest, and/or no phase agreement. But that did not happen. We find that the fluctuations in sea level are consistent with the biodiversity and rock package fluctuations. All of this agreement is additional evidence for the plausibility of a real, physical link between fluctuation in apparent diversity and the amount of preserved marine sediments, but it also suggests that sea-level per se is not the major driver because only one, minor part of the broad spectrum of sea-level changes is associated with the strong periodicity in diversity fluctuation.

## VII. Diversity and the geologic record: Bias or common cause?

*A. Interpretation regarding record quality and periodicity in fossil biodiversity*-- Given the concern that paleontologists have expressed over whether diversity patterns are strongly influenced by the variable quality of the geologic record (e.g. Peters and Foote 2001; Smith and McGowan, 2005, 2007) it is necessary to demonstrate that such an influence is not the whole story if one is to claim a biological component is involved in the periodicity in marine diversity, especially since, as we have seen, a 60 Myr periodicity is present in data on the number of sedimentary packages that make up the stratigraphic record of North America and a weak but consistent signal is also detectable in sea-level curves. Peters initially demonstrated a relationship between diversity and preserved sedimentary packages (Peters 2006a) but he did not discuss the  the periodicity of either. Peters considered the possible influence of record bias and concluded that, although some bias on diversity might be imposed by record quality,

"common cause" of real evolutionary turnover and record variation was also present.

But if a "record quality bias" were solely responsible for the periodicity in diversity what evidence should there be for its existence and does the record show it?

First, if the variation in record quality were the sole factor governing the periodic

fluctuation in SLG diversity in the marine realm one would expect the stratigraphic

record would carry a distinct periodic signal in several kinds of data: (a) sea-level; (b) a

regular appearance of global unconformities, and (c) variation in the volume of marine

sediments preserved. Sea-level curves do have a peak consistent with the diversity

periodicity, but it is simply one of many, none of which are dominant. Since diversity

only responds strongly to the 62 My periodicity among the various statistically minor but

equivalent sea-level periodicities it is more likely that diversity and this particular sea-

level peak are related through other factors rather than diversity variation being driven

by just this one and not all sea-level fluctuations. A periodic variation in the volume of

marine sediments does exist but, again, it is far more likely that this relates to "common

cause" (Peters 2006b, 2008a) rather than the periodicity in diversity simply being a

result of bias in the record. This is especially the case since, as just noted, no other

systematic variation in sea-level affects diversity systematically. Also, there is no

evidence of global gaps in the geologic record that would systematically affect diversity

data strongly.  If there were, it surely would have been detected in the intensive work

that has been done in calibrating and refining the absolute geologic time scale

(Gradstein et al. 2004a) and in compiling the extensive high-resolution correlation programs that have become an important part of modern stratigraphic analysis, such as graphic correlation and constrained optimization and similar techniques (Gradstein et al. 2004b; Harries 2003).

Secondly, the data used in the diversity curve developed by the Paleobiology Database (Alroy et al. 2008; Alroy 2008) has been designed and analyzed specifically to avoid sampling bias from the variable quality of both the geologic record and collecting intensity, so no periodicity should be displayed by it if the periodicity is produced by record bias, yet the periodicity signal is clearly present in the Paleobiology Database diversity curve (Melott 2008 and Paper I). If this data compiled to avoid record quality issues shows the periodicity, as it does, then record quality is unlikely to be the major factor in producing that periodicity.

Thirdly, the long-lived component of the marine fauna (LLG) displays very different evolutionary dynamics than the short-lived (SLG) component. If variation in geologic record quality were influencing the record of first appearances (by postponing the first appearance of genera that originated during times of poor record until the record quality improved) or last appearances (by truncating ranges to the time when the record becomes poor rather than the later time when extinction occurred during the time of poor record), then the clustering of those first and last appearances should be at the same times for both SLG and LLG. As far as first appearances (origination) of SLG and LLG are concerned, the patterns are *entirely* different and any peaking in LLGO is almost antiphase to SLGO (as discussed above in section IV). If variation in geologic record quality were influencing the record of first appearances (by postponing the first appearance of taxa that originated during times of poor record), the clustering of those first appearances should be at the same times for both SLG and LLG. The fact that this is not the case demonstrates that the patterns of origination observed are not produced by variation in record quality. SLG and LLG do share timing of elevated extinction (Fig. 6A), but analyses such as those by Peters (2006b, 2008a) indicate that true biotic turnover is associated with apparently elevated extinction rates even when correlated with lower amount of available record. In reference to gaps in the record as a source of clustering originations and extinctions Peters (2006b) says, "These results do not support the hypothesis that hiatuses at major unconformities alone have artificially clustered genus first and last occurrences, thereby causing many of the documented statistical similarities between the temporal structure of the sedimentary rock record and macroevolutionary patterns" (p. 387).

Because (a) diversity responds strongly to only one, not all periodic signals in sea-level, (b) data compiled is such a way as to avoid variable quality of the geologic record displays the observed 62 My periodicity, and (c) patterns of first appearances (origination) can not be related to variation in the quality of the geologic record we conclude that variation in record quality is not the predominant source of the periodicity observed in the interaction of SLG origination and extinction or in the timing of peaks in SLG origination, extinction and SLG diversity. Instead, as noted below, we support Peters' conclusions about common cause and feel the systematic fluctuation in diversity of SLG and the systematic fluctuation in number of marine sedimentary packages are also related through a shared causal system.

*B. Timing relation: sedimentary packages and mass extinctions.*--We note in Figure 12 that there is some disagreement between the biodiversity fits and the rock package fit near the present. The level of agreement is best near the time of the end-Permian extinction, which is just when all the biodiversity fits are closest, as in Figure 9A-C. Without prior knowledge of the apparent agreement of these two curves, one would probably use the concept of sampling bias — finding fewer fossils when there are fewer rock packages — and expect that the mass extinction events should occur when the rock package 62 Myr fit curve is below zero (the mid-point of the amplitude range of periodic fluctuation), as shown by the shaded regions in Figure 12. Our null hypothesis is that there is no such relationship, and only half should lie there. (Of course, this is a limited test of a continuous signal with a series of discrete points.) Testing this hypothesis, we find that 10 of the mass extinction events lie in the shaded zone, which is very close to the expectation value 9.5. Thus we cannot reject the hypothesis that mass extinction events are unrelated to the number of rock packages. This gives no support to the idea that mass extinctions are an artifact of sampling bias.

If we use the same approach as in section V. to examine the timing of 19 mass extinction events, i.e. examining the *rate of change* of Peters' rock package measure, a different answer results. 14 mass extinctions occur on the declining phase of rock packages, much as in the case of biodiversity (Fig. 12).  We can reject the hypothesis that mass extinctions are unrelated at the p=0.03 level. This is in agreement with the suggestion (Peters 2006b, 2008a) that extinctions and rock package numbers are results of some common cause, and not simply a preservation bias.  Note that mass extinctions are more often found when the number of rock packages is in decline, *even if currently still high.*

Additionally, it can be seen Figure 10C and 10D that the relationship between sedimentary packages and biodiversity is driven by changes over long time scales, not short-term ones.  Foote (2000) and Peters (2005) note that preservation bias would predict similarities driven by short, interval to interval variation in rates of recovery. Our results relate to long-term changes, not those that would be the result of short-term phenomena subject to such biases.

C. *The idea of common cause.*— Peters (2006b, 2008a) has demonstrated that genus extinction and faunal turnover relate closely to variation in the number of gap-bound marine sedimentary packages and the breaks between them, and effectively argues that while sampling bias may somewhat distort macroevolutionary patterns, the correspondence between patterns in the fossil and sedimentary records implies a biological signal as well, a linkage he refers to as "common cause." Episodes of faunal loss (extinction) and turnover (origination of new genera along with loss of old) correlate with times of less rock record. Thus breaks in the record and times of faunal turnover are both real and apparently share related (common) cause.

Since the observed 62 Myr periodicity in diversity is concentrated in the SLG and a similar periodicity is apparent in gap-bound sedimentary packages, especially those that are clearly marine in origin (carbonates), what might be the factors that link turnover in short-lived genera to variation in the amount of marine sedimentary record?

D. *Vulnerability of short-lived genera and common cause.*— Genus longevity can be influenced by (a) species richness, (b) geographic range, (c) local abundance, (d) life habit, (e) trophic strategy, (f) reproductive mode and (g) body size and inferred generation time (list from Jablonski 2005 and Liow 2007 and references therein). "Ecological versatility" (Liow 2007), a combination of extended geographic range, greater species richness and greater morphologic variability, is associated with increased genus longevity in ostracods and Powell (2007) has documented that increased geographic range size and reduced evolutionary volatility in brachiopods in the Carboniferous are closely related. For cohorts of Ordovician genera Miller (1997) demonstrated that as genera aged they were more likely to have representatives on several continents and in more environmental settings. Factors that enlarge the range of environments and/or locations in which representatives of a genus might live should make the genus more resistant to extinction because some species of the genus would be more likely to survive stresses that are locally or regionally concentrated than would be the case for genera with fewer species restricted in geographic range (more endemic) and/or environmental tolerance (more stenotopic). Since new (young) genera must begin as single species inhabiting an initial local area and it takes time to evolve multiple species that can be spread widely geographically or adapted to more diverse environments, it is reasonable to expect short-lived genera (SLG) to be more vulnerable to extinction than long-lived genera. Those genera that evolve more species that spread geographically or environmentally early in their histories are more likely to become LLG.

Because the genus data used for marine diversity analysis are predominantly collected from sedimentary rocks exposed on the continents, the times when fewer gap-bound sedimentary packages were deposited and preserved would be times that were also differentially hard on genera of restricted geographic or environmental range. Hence the parallelism of periodicity in diversity (evolutionary turnover) of SLG and Peters' data on gap-bound sedimentary packages, leads to our support for Peters' suggestion of common cause linking evolutionary turnover and the amount of marine sedimentary packages (Peters 2006b, 2008a). As Peters (2008b) has also noted, these features parallel both the hypothesis of extinction of perched faunas (Johnson 1974) and the influence of sea-level that Norman Newell once suggested might be the cause of major extinction events (Newell 1962, 1963 1967), albeit on a different temporal scale.

E. *Distribution of genus longevities.*— When the analyses were done for Paper I and this paper the only data available on genus longevities as a whole was the subdivision by Rohde and Muller (2005) into SLG and LLG with the average length of a period in the geologic time scale (45 My) as the criterion for assignment to short and long-lived categories. Our discovery of the difference in evolutionary dynamics of the two categories raises the question of whether this subdivision is at a fundamental point in the distribution of genus longevities or simply one of many possibilities. Until quite recently the distribution of genus longevities was not known, but Bornholdt *et al.* (2009) have now determined that the distribution of genus longevities in the Sepkoski genus database is a log-normal distribution and that, although each order has a different mean genus longevity (which is a major determinant of ordinal longevity), the distribution of genus longevities within orders is also log-normal. According to Bornholdt *et al.* (2009) the mean genus longevity is 27 Myr with a standard variation of 36.1 Myr. They also determined (Bornholdt *et al.* 2009: figures 6B and 6C) that ordinal longevity increases approximately as the square of average contained genus longevities up to an average genus longevity of 20 Myr and then flattens considerably for average genus longevities within orders greater than 20 Myr.

Because the distribution of genus longevities is continuous and scale-free from very short to very long (ranging from short single substages of less than two million years to longer than 200 million years) it is clear that the choice of 45 Myr as a break between SLG and LLG is arbitrary. However, in a continuum any twofold subdivision is arbitrary and the average length of a period in the geologic time scale is a reasonable choice, since it marks a length of time over which faunal turnover is sufficient to have generated global acceptance of the differences in faunas as marking these major time intervals. Smaller intervals of time (epochs and ages [series and stages]), while widely recognized, have been subject to extensive revision on many criteria, but the periods of the geologic time scale have been generally stable since most were first recognized in the nineteenth century. However, future research would be welcome on this topic. For example, it might be worth examining a range of longevity groupings to see how they fit the periodicity pattern. One comparison that would be of particular interest would be at 20 Myr, since that seems to be a point at which change in average genus longevity changes in its importance as related to ordinal longevity (Bornholdt *et al.* 2009). But for now, the subdivision used here seems as valid as any, and it is sufficient to recognize that significant differences in vulnerability and evolutionary dynamics characterize the SLG and LLG as designated by Rohde and Muller (2005) and analyzed in this paper.

### IX. Applicability to other published results

*A. Paleozoic versus post-Paleozoic evolutionary dynamics.*— Our results may account for previously unexplained changes in evolutionary dynamics reported by others (Foote 2000; Miller and Foote 2003, 2009).

Fig. 6C (see also Fig 4A, 4B) shows enhanced SLGE peaking approximately 15 Myr after SLGO. This tends to enhance fluctuations, but also negate any net long-term biodiversity change from enhanced SLGO. On the other hand, intervals of enhanced SLGE do not suppress or enhance subsequent SLGO. Consequently, diversity changes in the Paleozoic, when short-lived genera dominate (Fig. 1A) will be driven by extinction (Foote 2000). The change from Paleozoic to post-Paleozoic observed in evolutionary dynamics by Foote (2000), with a weak driving role for origination in the post-Paleozoic replacing the drive by extinction in the Paleozoic, is a consequence of the increased fraction of LLG in the marine fauna in the post-Paleozoic.  In Fig. 6D, LLGE is followed by enhanced LLGO; LLGO on the other hand is followed by suppressed LLGE. This combination would tend to make origination the stronger driver among LLG. The shift to the weak driving role of origination in the post-Paleozoic (Foote 2000) is consistent with the LLG becoming the consistent majority only during and after the Jurassic, and even then not yet fully dominating fluctuation in total diversity.

In the post-Paleozoic, cohorts originating in intervals immediately following major mass extinctions include a high proportion of longer-lived genera, although this was not the case during the Paleozoic (Miller and Foote 2003). This pattern may relate to enhanced LLGO following LLGE (Fig 6D). We know that LLGE and SLGE largely coincide (Fig 6A), but SLGO does not increase following SLGE (Fig 6C). LLG, with their enhanced origination following extinction, became consistently numerically dominant at and after mass extinctions only in the post-Paleozoic.

In another report Miller and Foote (2009) note that origination rates of genera restricted to ocean-facing areas became consistently greater than those restricted to epicontinental seas only during the Jurassic and after. .Although it would take a major research effort to prove the relationship, it seems highly likely that the observation by Miller and Foote (2009) about continuous higher origination rates in ocean-facing ecosystems beginning in the Jurassic may be related to the rise of LLG to diversity dominance. We know that the increase in global diversity in the post-Paleozoic is predominantly an increase in LLG. It is reasonable to expect the global increase to be occurring in both ocean-facing and epicontinental settings and that LLG should account for most of that increase in both settings. Because of the higher physiological stresses for life in epicontinental settings (where salinity and temperature variability are likely to be irregular and potentially large compared to the very stable salinity regime and less likely extremes of temperature expected in open-ocean settings, making epicontinental seas generally less "equable" than open ocean settings) we would predict that epicontinental seas would have had a higher proportion of "resistant" genera than "vulnerable" genera, whereas open-ocean settings should have both "vulnerable" and "resistant" genera. We believe "vulnerable" and "resistant" genera are, in large measure, what we see as SLG and LLG. The evolutionary radiation of LLG in the post-Paleozoic should be particularly noticeable in open-ocean facing settings because the amount of epicontinental seas decreased drastically, especially in the Cenozoic, so as post-Paleozoic time passed less and less of total diversity can be from the fauna restricted to epicontinental settings. Therefore the coincidence of timing of the Miller and Foote (2009) observation about change in origination in ocean-facing settings and the timing when the rise in diversity of LLG becomes prominent suggests a connection to the results of our study.

In all three of these cases the original observations dealt only with the total fauna and no causes were identified. It seems that no differences in global physical conditions affecting evolution or general changes in evolutionary dynamics need be involved. The actual basis for the phenomena observed simply may be the change in diversity dominance from SLG to LLG during the Mesozoic, with each group retaining its particular pattern of evolutionary dynamics unchanged, but with the differences in evolutionary volatility of SLG versus LLG making the transition inevitable.

*B. Extinction-origination: Our stars or ourselves?*--We have discussed a number of choices in the way data are processed and noted that most of them do not qualitatively affect the sense of results.  Now we present one choice which makes a great deal of difference, and show how our results are consistent with a previous result which may at first glance appear to be inconsistent.  It concerns the cross-correlation of extinction with origination (Fig. 6).

In Kirchner and Weil (2000), the correlations were based on the proportion of genera (either as a count per stage, or as a rate per Myr).  If we repeat this computation, we get results consistent with theirs, and opposite to what we found with the SLG (compare Fig 13 with Fig 6C). This is true whether we examine rates per Myr or per stage. Purdy (2008) found results consistent with ours but was not able to explain his discrepancy with Kirchner and Weil (2000). This can be understood as follows: Because the diversity of SLG changes so much during diversity cycles there are strong differences in the proportional importance of individual SLG genera at different times in each cycle. Therefore, following major diversity decrease, the total diversity is reduced, and any O shows an elevated proportion. Following major diversity increase, elevated diversity depresses the proportion of any E. The Kirchner and Weil result is "actuarial", based on estimated probabilities of the fate of genera; our result is based on the numbers of events and thus to their sums which produce the changes in diversity recorded in the periodicity.

We have found that such cross-correlations do not change a great deal between the Paleozoic and the post-Paleozoic, but we do find consistent differences of exactly this kind for all classes and cuts: using proportional changes, the result at positive lag is increased and the result at negative lag is decreased (made more negative). Positive lag means extinction leading; when there is substantial extinction the number of genera is decreased. Subsequently any origination is proportionately greater, because fewer genera are available to give rise to new genera.  Similarly when extinctions lag an interval of increased origination, the extinction is proportionately smaller—the origin of the decrease at negative lag. Kirchner and Weil (2000) is relevant if one is to take odds on the fate of a genus based on the previous few 10 Myrs; ours are more relevant to constructing the dynamics of diversity.  It is changes in the numbers of genera, not the proportions of genera, that sum to produce changes in diversity over time.

## X. Summary and Conclusions

A. *Summary of results.*— Our investigations can be summarized as:

1. The interactions of SLGO and SLGE rates together, not fixed behavior of either alone, is responsible for producing the 62 Myr periodicity of fluctuation in diversity that is ubiquitous in data on the marine fauna.

2. Although LLG also respond to major extinction events, LLGO is unrelated to SLGO and the overall evolutionary behavior of LLG is different than that of SLG.

3. Although the full set of 62 extinction events, 60 listed in Barnes et al. (1996) plus two others, is evenly distributed between the phases of increasing and decreasing diversity comprising the 62 ± 3 Myr periodicity, the vast majority (15 of 19) of the mass extinctions identified by Bambach (2006) fall in the decreasing diversity phases. Applying the Arens and West (2008) concept of "press-pulse" it may be that equivalent stresses (pulses) during more equable intervals (times of less "press") might fail to produce the magnitude and rate of extinction produced during less equable times, which could account for the asymmetries in the distribution of both mass extinctions and less severe extinction events.

4. A 58 ± 4 Myr periodicity in the number of preserved gap-bound sedimentary packages, especially distinct for marine carbonage packages, is nearly In phase with the 62 ± 3 Myr periodicity observed in marine diversity.

5. Although a periodicity in sea-level similar to that for marine diversity and gap-bound sedimentary packages is not obvious or statistically important, a weak periodicity at 57 Myr with the appropriate timing does exist.

6. The 62 ± 3 Myr periodicity and its associated evolutionary dynamics are not an artifact of variable quality of the geologic record, as demonstrated by the presence of the periodicity in the PBDB data, which was compiled to avoid sampling biases, and is further attested to by the differences in the record of origination of LLG and SLG and other factors. The association between periodic diversity change and periodic variation in the number of gap-bound sedimentary packages relates to what Peters (2006b, 2008) has called "common cause."

7. Patterns of genus longevity observed in a variety of paleontological studies (Jablonski 2005, Liow 2007, Miller 1997 and references therein) indicate that SLG are probably genera that are less species rich and more restricted in geographic range and/or environmental tolerance than LLG. These characteristics would make SLG more vulnerable to extinction than LLG.

8. The difference in evolutionary dynamics between SLG and LLG, which makes SLG evolutionarily more volatile, explains why the proportion of LLG in the total fauna has increased over time while SLG diversity has risen and fallen but not had an overall trend toward increasing diversity.

9. The differences in general evolutionary dynamics between the Paleozoic and post-Paleozoic, observed in several studies by others, are probably related to the differences in evolutionary dynamics of SLG compared to LLG, which remain constant over time, and to the rise to diversity dominance by LLG during the mid-Mesozoic. No change in biological or physical processes is needed to explain the differences in evolutionary dynamics of the total fauna between the Paleozoic and post-Paleozoic.

B. *Conclusions.*— What conclusions can we draw from these results? Apparently there is a systematic fluctuation in environmental conditions in the world's oceans that drives periodic fluctuation in diversity. Two potential causal systems are known, one related to the possible effects of periodic sea-level change as recorded in the number of gap-bound sedimentary packages and also weakly expressed in sea-level curves. The other is astronomical.

The "common-cause" approach would ascribe the pattern of diversity fluctuation of SLG to the waxing and waning of habitat stresses and opportunities for the evolution of new genera as the area and stability of shallow seas changed on the continents. But what would make this pattern so regularly periodic? If the 62 Myr periodicity is driven by major shifts in the degree of continental flooding, why is the 62 Myr periodicity so weak as to be buried and insignificant in both sets of sea-level data? The marine deposits (both carbonates and clastics) from which most fossils have been collected formed when sea-level relative to the surface of the continent was high enough that the sea could flood widely onto the continental surface, so the variation in numbers of gap-bound sedimentary packages clearly must relate to continental freeboard in a way that is not captured by the sea-level data. The range of sea-level change is highly variable, but it still should be possible to find any particular long-term periodicity pattern if such a frequency makes a strong contribution when compared with all other periodicities present in sea-level data. There are clearly time periods both shorter and longer than the ~60 Myr signal that strongly contribute to the power spectrum of sea level data (Fig. 11A), but not to those for sedimentary packages (Fig. 10A through 10D) or biodiversity (see Paper I). However, it is just the 62 Myr periodicity that is strong for biodiversity and in the variation in preserved sedimentary packages.

Tectonic processes (uplift and subsidence) are the processes that operate at the appropriate time scales. Large regional effects that could affect SLG but not have much influence on more widespread LLG can be tectonically produced. Change in environmental equability on a long-term basis, through such things as climate change from orographic effects, also could be a function of major tectonic episodes. But no periodic patterns of internal earth behavior that might be regarded as a "pulse of the Earth" have yet been convincingly demonstrated. We suggest that investigation of possibilities for periodic tectonic influence on continental freeboard is needed.

An alternate hypothesis which naturally gives rise to such regularity is a possible astronomical effect. Astronomical orbits are typically gravitational processes which are often nearly dissipationless and therefore may remain periodic for long times as with the ~100 Myr timescales typical of galactic dynamics. For example, the low points of the 62 Myr periodicity cycles in biodiversity correspond in timing with excursions of the Solar System to Galactic north, a nearly periodic 63.6 Myr oscillation. It has been hypothesized that increased cosmic ray flux may be characteristic of Galactic north because of a possible shock wave produced by the movement direction of the Galaxy and that the low point in biodiversity might be a result of the effects of greatly increased exposure to high-energy cosmic rays (*TeV* to *PeV*) (Medvedev and Melott 2007; Melott et al. 2010). Such irradiation could possibly affect cloud formation but only on long timescales (Lockwood and Fröhlich 2007; Atri et al. 2010b); it may make large changes in the amount of lightning (Erlykin and Wolfendale 2010). Muons and thermal neutrons on the ground may produce DNA damage and damage to proteins (Turner 2005), providing a long-term stress which may interact with other environmental factors in a way that impacts biodiversity. Although these effects, if present, can account for the timing of the periodicity, it is not clear why the periodicity is expressed in marine diversity and not in terrestrial diversity.

A possible driver is that the climate is altered by the ionization, cloud cover, and lightning rate changes mentioned above. There is intense debate over the relative contributions of tectonics and erosion (Molnar and England 1990; Egholm et al. 2009; Dowdeswell et al. 2010), but effects of increased erosion and subsequent uplift due to unloading will affect continental freeboard. Work in progress (Melott, Bambach, and Petersen 2011) should shed further light on this possibility. We stress that this is a physically reasonable but speculative connection motivated by the fact that the period and phase of the Galactic oscillation is very similar to that observed in biodiversity (Medvedev and Melott 2007).

What does seem clear is (a) that a 62 Myr periodicity in marine diversity is unambiguous, (b) it is related to the regular fluctuation in diversity of SLG and (c) that the SLG are behaving as "canaries in the coal mine" with regard to some as yet undetermined driver for periodic fluctuation in environmental equability. Long-term tectonic patterns that can influence the destruction of endemic faunas plus also causing possible alteration of earth surface conditions like climate and climatic variability that could affect regional or global environmental equability may be being signaled by these paleontolgical patterns. Likewise, phenomena connected to astronomically related variation (Medvedev and Melott 2007; Melott and Thomas 2011) can, in principle, influence the equability of the earth's surface habitats (radiation, climate influences, etc.) and could be involved in altering the chances of nascent genera vulnerable to extinction before becoming widely distributed or that remain vulnerable by failing to expand in geographic or environmental range. In either case, SLG would gain in diversity as equability improved, both from increasing success of origination and decreasing extinction, and SLG would lose diversity as equability decreased, causing extinction of these more vulnerable genera and also decreasing the likelihood of new vulnerable forms becoming established, therefore decreasing apparent origination. Both tectonic and astronomical causal mechanisms are primarily hypothetical at this point, but both need investigation. The 62 Myr periodicity of biodiversity tells us that some highly regular periodic process influences marine life on earth. That process or processes needs explanation.

## Acknowledgments

We thank J. Alroy, P. Flemings, K. Miller, J. Ogg and S. Peters for providing some of the data which we analyze herein. T. Hallam made helpful comments. We are grateful to the American Astronomical Society for the sponsorship of the 2007 Honolulu multidisciplinary Splinter Meeting at which discussions leading to this project took place. A question from Michael Foote (at our presentation at the North American Paleontological Convention in 2009) about how origination and extinction behave as marine diversity periodically fluctuates focused us on this topic. Research support at the University of Kansas was provided by NASA Program Astrobiology: Exobiology and Evolutionary Biology under grant number NNX09AM85G. This paper is PBDB publication Number XXX.

## Literature Cited

Alroy, J. 2008. Colloquium Paper: Dynamics of origination and extinction in the marine fossil record. Proceedings of the National Academy of Science www.pnas.org/cgi/doi/10.1073/pnas.0802597105.

Alroy, J. et al. 2008. Phanerozoic trends in the global diversity of marine invertebrates. Science 321:97-100.

Arens N. A., and I.D. West 2008. Press-Pulse: A general theory of mass extinction? Paleobiology 34:456-471.

Atri, D., A.L. Melott, and B.C. Thomas. 2010a. Lookup tables to compute high energy cosmic ray induced atmospheric ionization and changes in atmospheric chemistry. Journal of Cosmology and Astroparticle Physics JCAP05(2010)008 doi: 10.1088/1475-7516/2010/05/008.

Atri, D., B.C. Thomas, and A.L. Melott. 2010b. Can periodicity in low altitude cloud cover be induced by cosmic ray variability in the extragalactic shock model? Astrobiology, submitted. arXiv:1006.3797

Bambach, R. K. 2006. Phanerozoic biodiversity mass extinctions. Annual Review of Earth and Planetary Sciences 34:127-155.

Barnes, C., A. Hallam, D. Kaljo, E. G. Kauffman, and O. Walliser. 1996. Global event stratigraphy. Pp. 319–333 in Walliser, O. (Ed.), Global Events and Event Stratigraphy In the Phanerozoic. Springer, Berlin.

Benton, M. J. 1993. *The Fossil Record 2*. (Chapman & Hall, London) 845 pp.

___. 1995 Diversification and extinction in the history of life.  Science 268:52-58. DOI: 10.1126/science.7701342

Bornholdt, S., K. Sneppen, and H. Westphal. 2009. Longevity of orders is related to the longevity of their constituent genera rather than genus richness. Theory in Biosciences 128:75-83.

Cornette J. L. 2007. Gauss-Vaníček and Fourier Transform Spectral Analyses of Marine Diversity.  Computing in Science and Engineering 9:61-63.

Dowdeswell J. A., D. Ottesen, and L. Rise 2010. Rates of sediment delivery from the Fennoscandian Ice Sheet through an ice age. Geology 38:3-6. doi: 10.1130/G25523.1.

Egholm D.L., S.B. Nielsen, V. K. Pedersen, and J. E. Lesemann 2009. Glacial effects limiting mountain height. Nature 460:884-888.

Erlykin, A.D., and A.W. Wolfendale (2010) Long term variability of cosmic rays and possible relevance to the development of life on Earth. To be published in Surveys Geophys. arXiv:1003.0082

Foote M. 2000. Origination and extinction components of taxonomic diversity: Paleozoic and post-Paleozoic dynamics. Paleobiology 26:578-605.

Foote M. 2005. Pulsed origination and extinction in the marine realm. Paleobiology 31:6-20.

Gradstein, F. J., G. Ogg, and A. G. Smith. 2004a. A Geologic Time Scale 2004.

Gradstein FM, Coope RA, Sadler PM (2004b) Biostratigraphy: time scales from graphic and quantitative methods. Pp. 49–54 in Gradstein, F. M., Ogg, J. G., and Smith, A. G. (eds.) A Geologic Time Scale 2004. (Cambridge University Press, Cambridge, UK).Cambridge University Press, NY.

Hallam, A., and P. B. Wignall. 1997. Mass Extinctions and Their Aftermath. Oxford University Press, Oxford.

Haq, B., and A.M. Al-Qahtani. 2005. Phanerozoic cycles of sea-level change on the Arabian Platform. GeoArabia 210:127-160.

Haq, B., J. Hardenbol, and P.R. Vail. 1987. Chronology of Fluctuating Sea Levels since the Triassic. Science 235:1156 -1167.

Haq, B., and S.R. Schutter. 2008. A Chronology of Paleozoic Sea Level Changes. Science 322:65-68

Hardenbol, J., T. Jacquin, and P.R. Vail (Eds.), 1998. Mesozoic and Cenozoic Sequence Stratigraphy of European basins. SEPM Spec.Publ.60:1-786.

Harries, P.J. (ed.) (2003) High-Resolution Approaches in Stratigraphic Paleontology. *Topics in Geobiology 21*. (Kluwer Academic Publishers, Boston).

Jablonski, D. 1986. Background and mass extinctions: the alternation of macroevolutionary regimes. Science 231:129–133.

___. 1989. The biology of mass extinction: a palaeontological view. Philosphical Transactions of the Royal Society of London, Series B, Biological Sciences 325:357–368.

—. 2005. Mass extinctions and macroevolution. Pp. 192–210 in Vrba, E. S. and Eldredge, N. (eds.), *Macroevolution: Diversity, Disparity, Contingency*. Supplement to *Paleobiology* 31:2.

Johnson, J. G. 1974. Extinction of perched faunas. Geology 2:479–482.

Kirchner, J.W. and A. Weil 2000. Delayed biological recovery from extinctions throughout the fossil record. Nature 404:177-179.

Krug, A.Z., Jablonski, D., and Valentine, J.W. 2009. Signature of the End-Cretaceous Mass Extinction in the Modern Biota. Science 323:767-771. DOI: 10.1126/science.1164905

Lieberman, B. S. and A. L. Melott. 2007. Considering the Case for Biodiversity Cycles: Re-Examining the Evidence for Periodicity in the Fossil Record. PLoS One doi: 10.1371/journal.pone.0000759.

Lockwood M. and C. Fröhlich 2007. Recent oppositely directed trends in solar climate forcings and the global mean surface air temperature. Proceedings of the Royal Society A doi:10.1098/rspa.2007.1880.

Lugowski, A. and J. Ogg 2010. TimeScale Creator (version 4.2.1) and data sets available at https://engineering.purdue.edu/Stratigraphy/tscreator/index/index.php.

Medvedev, M. V. and A. L. Melott 2007. Do extragalactic cosmic rays induce cycles in fossil diversity? Astrophysical Journal 664: 879-889.

Melott, A .L. 2008. Long-term cycles in the history of life: Periodic biodiversity in the Paleobiology Database. PLoS ONE  arXiv:0807.4729

Melott, A .L., and R. K. Bambach 2011. A ubiquitous ~62-Myr periodic fluctuation superimposed on general trends in fossil biodiversity. I. Documentation. Paleobiology, 37:92-112 (in press).

Melott, A.L. and R.K. Bambach 2010. Nemesis Reconsidered. Monthly Notices of the Royal Astronomical Society Letters, 407:L99-L102.

Melott, A .L., B. C. Thomas, D. P. Hogan, L. M. Ejzak, and C ,H. Jackman. 2005. Climatic and Biogeochemical Effects of a Galactic Gamma-Ray Burst.  Geophysical Research Letters 32: L14808 doi:10.1029/2005GL023073

Melott, A .L, D. Atri, B. C. Thomas, M. V. Medvedev,  G. W. Wilson, and M.J. Murray. 2010. Atmospheric consequences of cosmic ray variability in the extragalactic shock model II: Revised ionization levels and their consequences. Journal of Geophysical Research—Planets 115:E08002  doi:10.1029/2010JE003591.

Melott, A. L. and B. C. Thomas. 2009. Late Ordovician geographic patterns of extinction compared with simulations of astrophysical ionizing radiation damage Paleobiology 35:311-320.

Melott, A.L., and B.C. Thomas. 2011. Astrophysical Ionizing Radiation and the Earth: A Brief Review and Census of Sources. Astrobiology, submitted.

Miller, A. I. 1997. A new look at age and area: the geographic and environmental expansion of genera during the Ordovician radiation. Paleobiology 23:410–419.

Miller A.I. and M. Foote 2003. Increased Longevities of Post-Paleozoic Marine Genera After Mass Extinctions. Science 302:1030-1032.

Miller A.I. and M. Foote 2009. Epicontinental Seas Versus Open-Ocean Settings: The Kinetics of Mass Extinction and Origination. Science 326:1106-1108.

Miller, K.G., M.A. Kominz, J.V. Browning, J.D. Wright, G.S. Mountain, M.E. Katz, P.J. Sugarman, B.S. Cramer, N.Christie-Blick, and S.F. Pekar. 2005. The Phanerozoic Record of Global Sea-Level Change. Science 310:1293—1298. DOI: 10.1126/science.1116412

Molnar, P., and P. England 1990. Late Cenozoic uplift of mountain ranges and global climate change: chicken or egg? Nature 346:29-34.

Newell, Norman D. 1962. Paleontological gaps and geochronology. Journal of Paleontology 36:592–610.

___. 1963. Crises in the history of life. Scientific American 208:2:76–92.

___. 1967 Revolutions in the history of life. Special Paper of the Geological Society of America 89:63-91.

Peters, S.E., and M. Foote 2001. Biodiversity in the Phanerozoic: a reinterpretation. Paleobiology 27:583–601.

Peters, S. E. 2005. Geologic constraints on the macroevolutionary history of marine animals. Proceedings National Academy of Science 102:12326-12331.

___. 2006a. Macrostratigraphy of North America. Journal of Geology 114:391–412.

___. 2006b. Genus extinction, origination, and the durations of sedimentary hiatuses. Paleobiology 32:387–407.

___. 2008a. Environmental determinants of extinction selectivity in the fossil record. Nature doi:10.1038/nature07032.

—. 2008b. Macrostratigraphy and its promise for paleobiology. Pp. 205-232 in P.H. Kelley and R.K. Bambach, eds. From evolution to geobiology: research questions driving paleontology at the start of a new century. The Paleontological Society Papers, Vol. 14.

Powell, M. G. 2007. Geographic range and genus longevity of late Paleozoic brachiopods. Paleobiology 33:530–546.

Purdy, E.G. 2008. Comparison of taxonomic diversity, strontium isotope, and sea-level patterns. International Journal of Earth Sciences 97:651-664.

Raup, D. M. and Sepkoski, J. J. Jr. 1984. Periodicity of extinctions in the geologic past. Proceeding of the National Academy of Sciences, U. S. A. 81:801–805.

Rohde, R. A. and R. A. Muller 2005.  Cycles in fossil diversity. Nature 434: 208-210.

Scargle, J. D. 1982 Studies in astronomical time series analysis. II. Statistical aspects of spectral analysis of unevenly spaced data.  Astrophysical Journal 263:835-853.

Sepkoski. J. J., Jr. 1996. Patterns of Phanerozoic extinction: a perspective from global data bases. Pp. 35–51 in Walliser, O. (Ed.), Global Events and Event Stratigraphy In the Phanerozoic. Springer, Berlin.

—. 2002. A compendium of fossil marine animal genera. Bulletin of American Paleontology 363, eds Jablonski, D., Foote, M.  Paleontological Research Institution, Ithaca, NY.

Smith, A B., and A. J.McGowan. 2005. Cyclicity in the fossil record mirrors rock outcrop area. Biological Letters 1:443–445. doi: 10.1098/rebl.2005.0345.

Smith A.B.,and A.J. McGowan. 2007. The shape of the Phanerozoic marine Palaeodiversity curve: how much can be predicted from the sedimentary rock record of western Europe? Palaeontology 50:765–774.

Turner, J. 2005. Interaction of ionizing radiation with matter. Health Physics 88:520-544.

Wilgus, C. K., C. A. Ross, and H. Posamentier (eds.) 1988. Sea-Level Changes: An Integrated Approach (Society of Economic Paleontologists, Mineralogists Special Publication No. 42) data downloaded from

http://hydro.geosc.psu.edu/Sed_html/strata_front_page.html

## Figure Captions

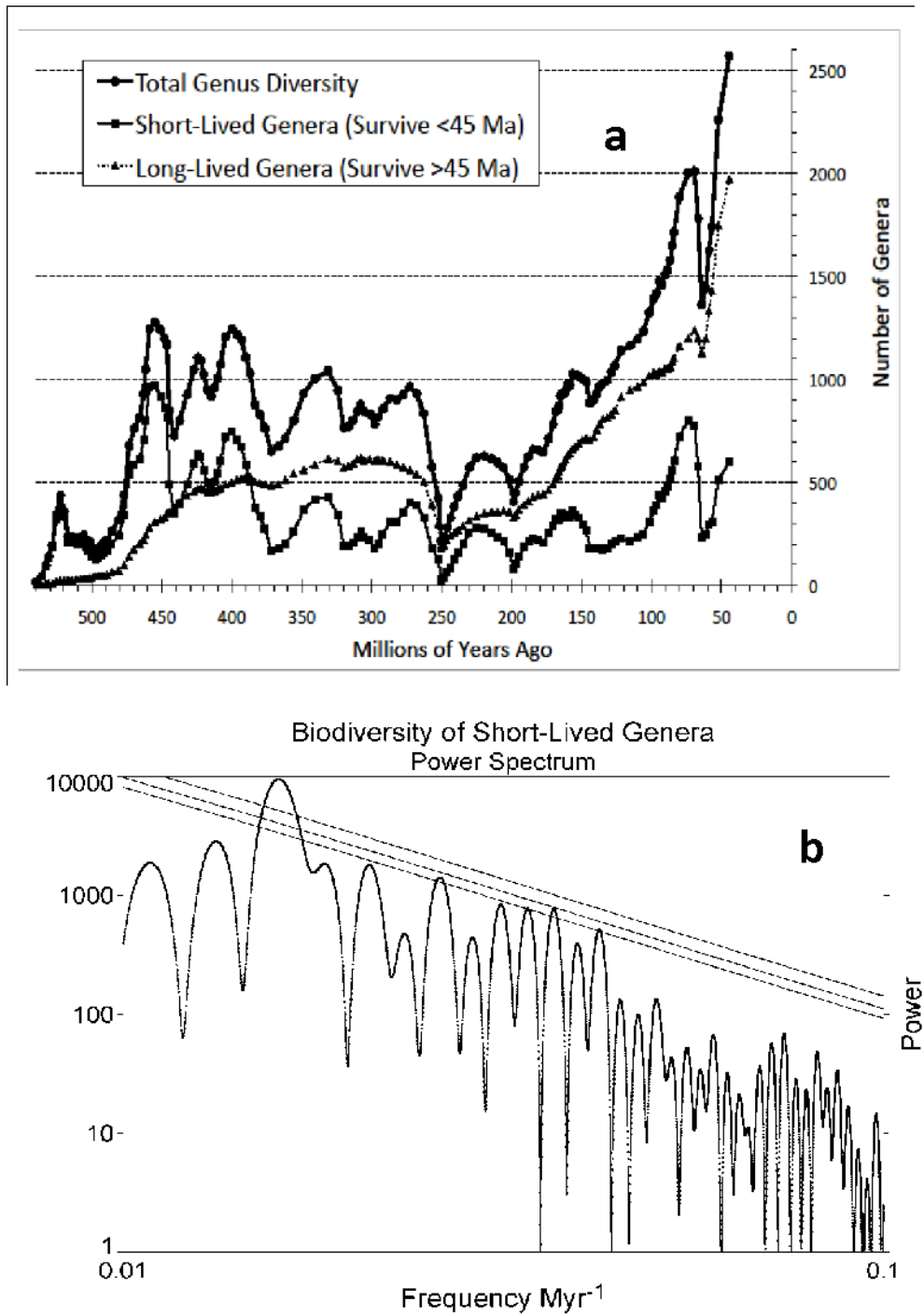

Figure 1 (A) The history of marine genus diversity and its fluctuations. The number of well-resolved genera is the sum of long-lived (>45 Myr) (LLG in text) and short-lived

(SLG in text). The long-lived genera (middle line) are characterized by a steady increase with small fluctuations except at two major mass extinctions. The short-lived genera do not show much increase over time, and show strong fluctuations. (B) A log-log plot of the power spectrum of fluctuations in detrended short-lived genera against frequency f=1/T.  As in other logarithmically scaled plots herein, the axis labels reflect the values, not their logarithms. Confidence intervals are at p=0.05, 0.01, and 0.001 against such a peak arising anywhere in the spectrum. The obvious peak at T=62 Myr, which is the one of interest for this paper, exceeds the ability of our software to resolve the p-value. There is also potentially interesting structure in the several peaks in the center that reach lower levels of significance in the vicinity of 30 Myr which we do not consider at this time.

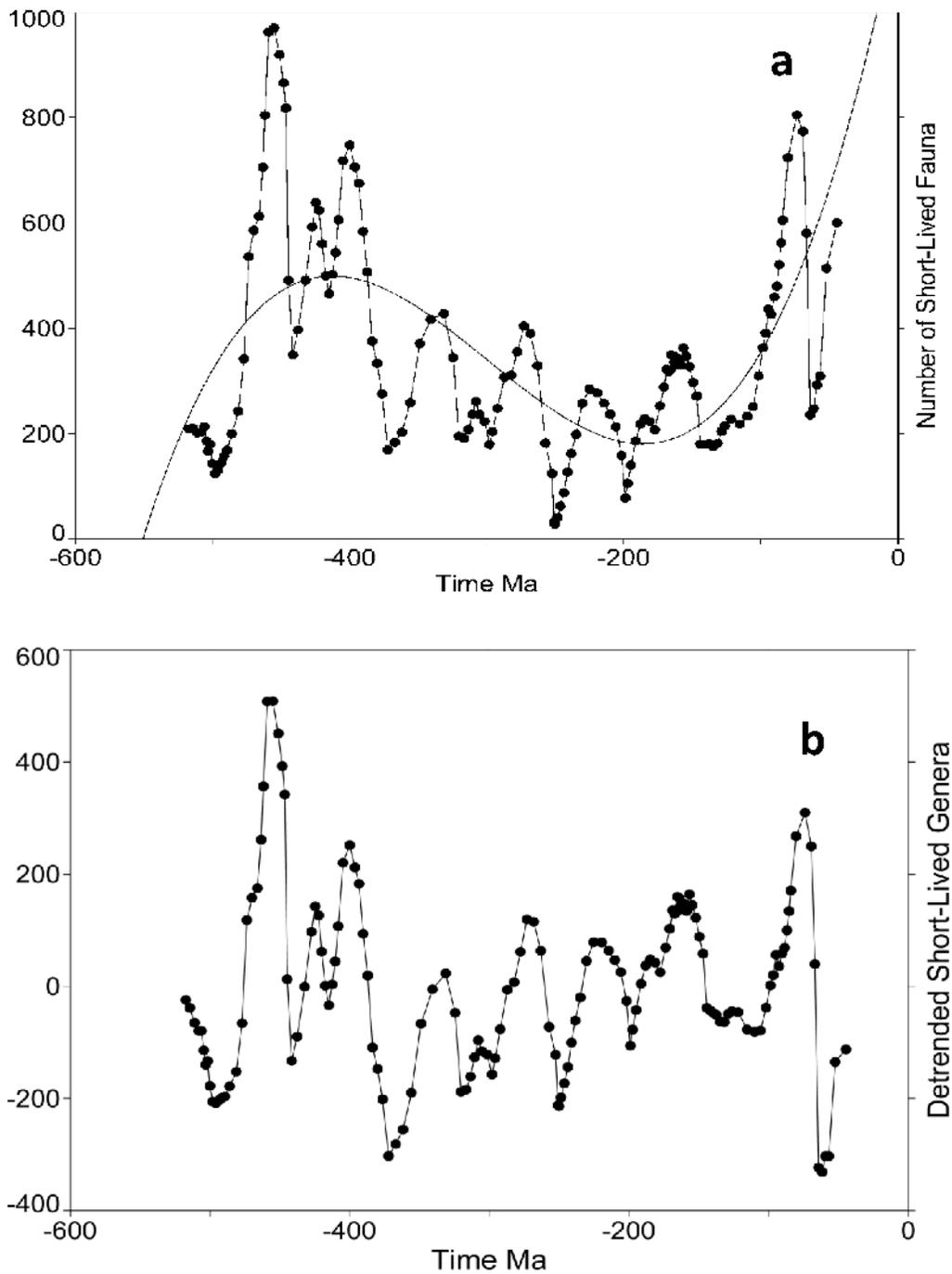

Figure

Figure 2: Detrending of the time series of the number of short-lived genera, as an example of the procedure. (A) A cubic polynomial is fit to the data by least-squares minimization. (B) The cubic fit is subtracted from the data, eliminating long-term trends.

The residuals are plotted here, and are the basis for all time series analysis in this study. Note that the values are both positive and negative depending on whether the actual diversity is greater or less than the cubic polynomial trend line.

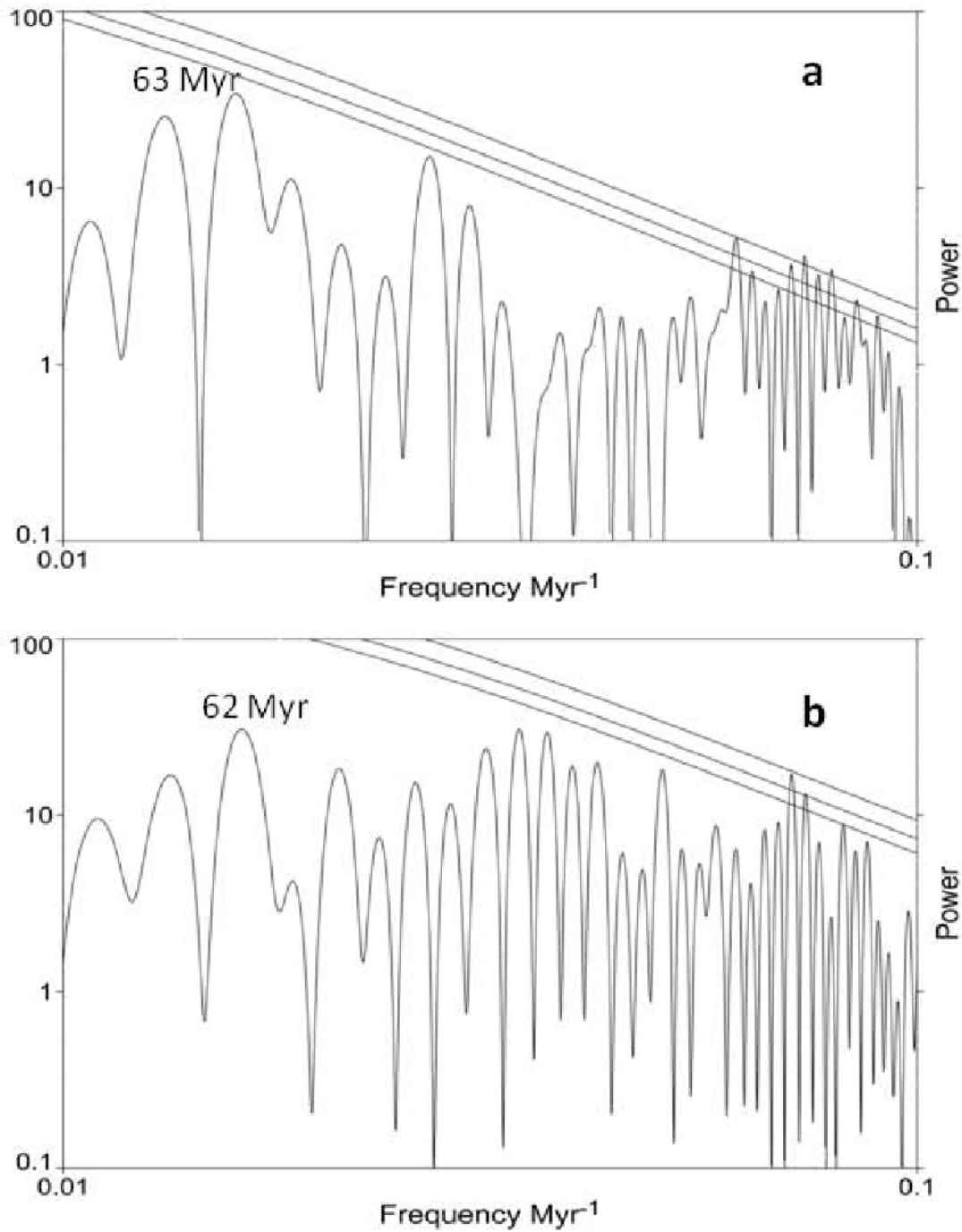

Figure 3: (A)  A log-log scaled plot of the power spectra of the detrended time series for fluctuations in origination rate (per Myr) of short-lived genera. As before, axis labels indicate the values, not their logarithms. The parallel lines in this figure and all other power spectra illustrated in this paper indicate values of p=0.05, 0.01, and 0.001 against the any spectral peak of that height given a Gaussian distribution. These rates show increased high-frequency noise compared with the raw biodiversity spectrum seen in Fig. 1B.  At this time we do not discuss a few peaks at shorter periods which appear to be significant, while not contributing a large amount to variance.  (B) As in 3A, for extinction rate of short-lived genera.

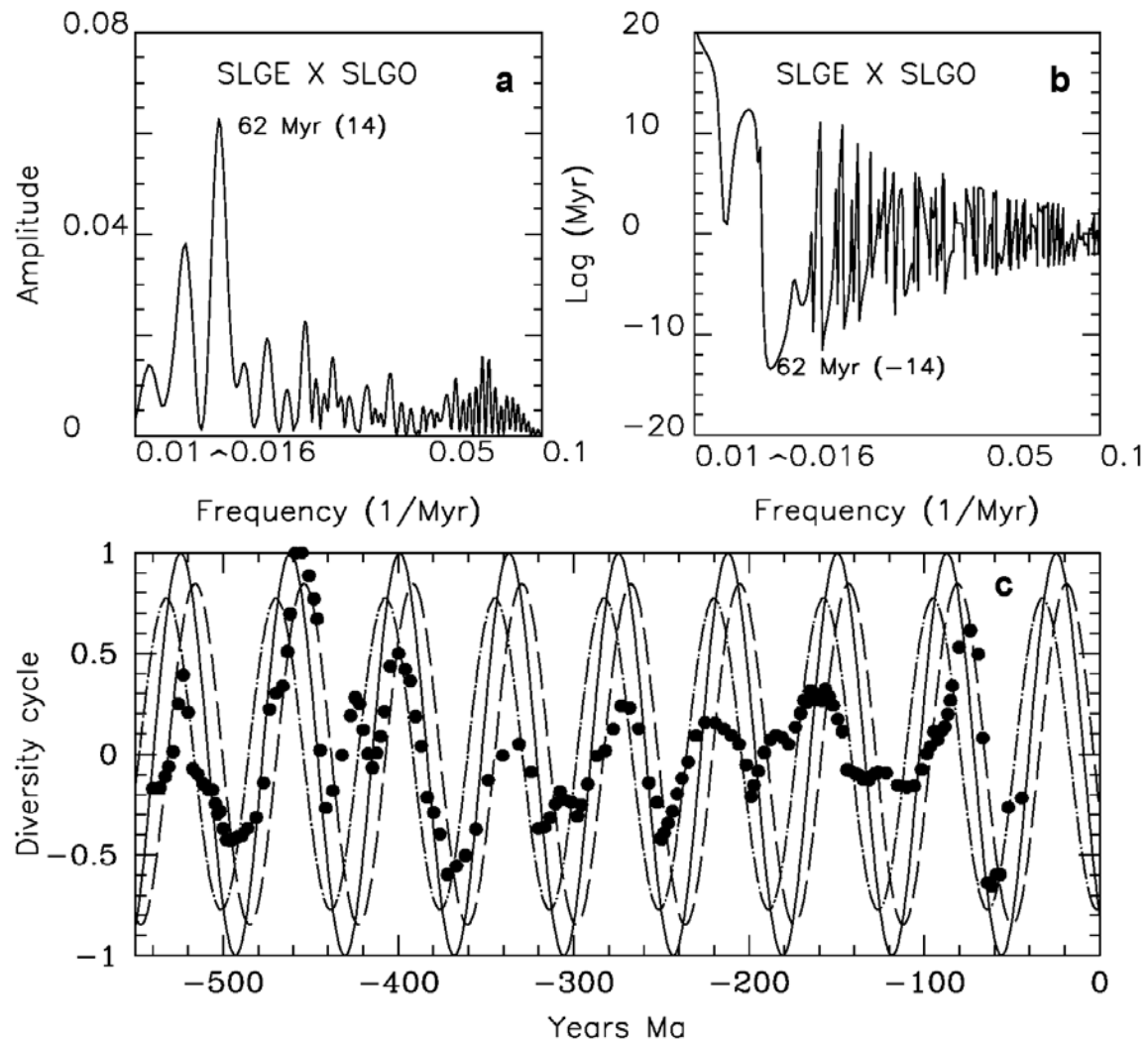

Figure 4: Characterization of the origin of the fluctuations in Fig. 1 in origination and extinction of only short-lived genera. (A) The amplitude of the normed (complex) cross-spectrum of the extinction and origination rate against frequency, scaled logarithmically. This quantity measures the joint amplitude of the two spectra without regard to their phase, i.e. without regard to whether the peaks at a given frequency occur at the same time. A peak at 62 Myr strongly dominates; inspection of the distribution of phases indicates a 14 Myr phase lag at the peak, with origination leading. (B) A plot of phase lags in Myr against log f. At periods longer than the 62 Myr dominant, the lag is positive.

At periods shorter than 62 Myr, it is not coherent. There is a wide band of negative lag visible in B around the peak seen in A, so that the whole dominant band contributes with extinction lagging origination. (C) Sine waves show the position of the best-fit to the spectral peak in short-lived genus diversity (solid line), origination rate (dash-dot line), and extinction rate (dashed line). The filled circles are the deviations of detrended diversity values from the trend line as described in Fig. 2, and divided by the maximum value.

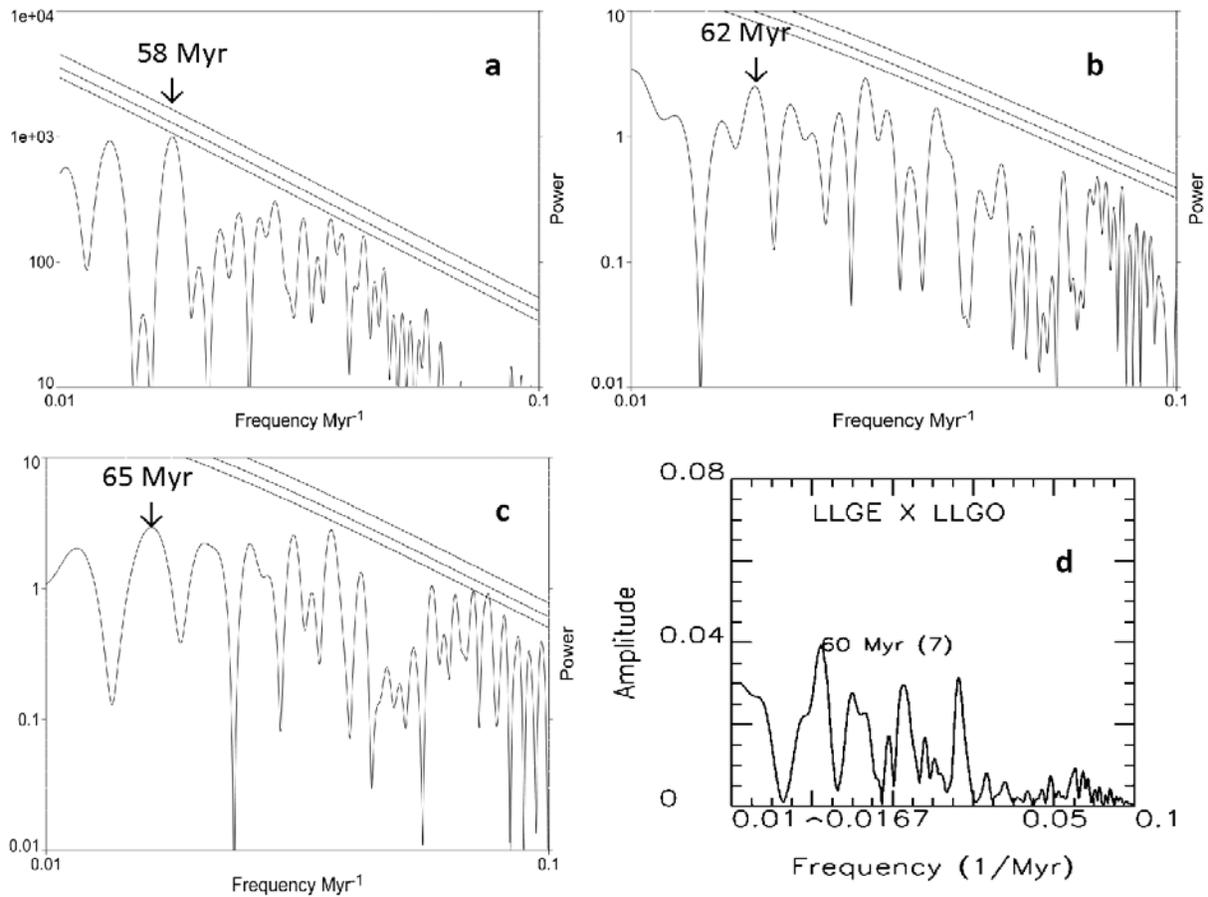

Figure 5: Spectral behavior of the long-lived genera. (A) The power spectrum of LLG diversity fluctuations. The overall amplitude is lower than that seen for short-lived genera in Fig. 1B, consistent with the generally lower level of fluctuations. There is a peak at 58 Myr, but it is not significant. (B and C) The power spectra of LLG origination rate and extinction rate, respectively. Neither contains a peak significant above the background, but both have peaks equal within the errors to 62 Myr. (D) The complex amplitude cross-spectrum of long-lived genera extinction and origination as a function of log frequency shows a peak at 60 Myr; it is the largest amplitude peak within the range of interest, but does not dominate as does the peak in Fig. 4A.

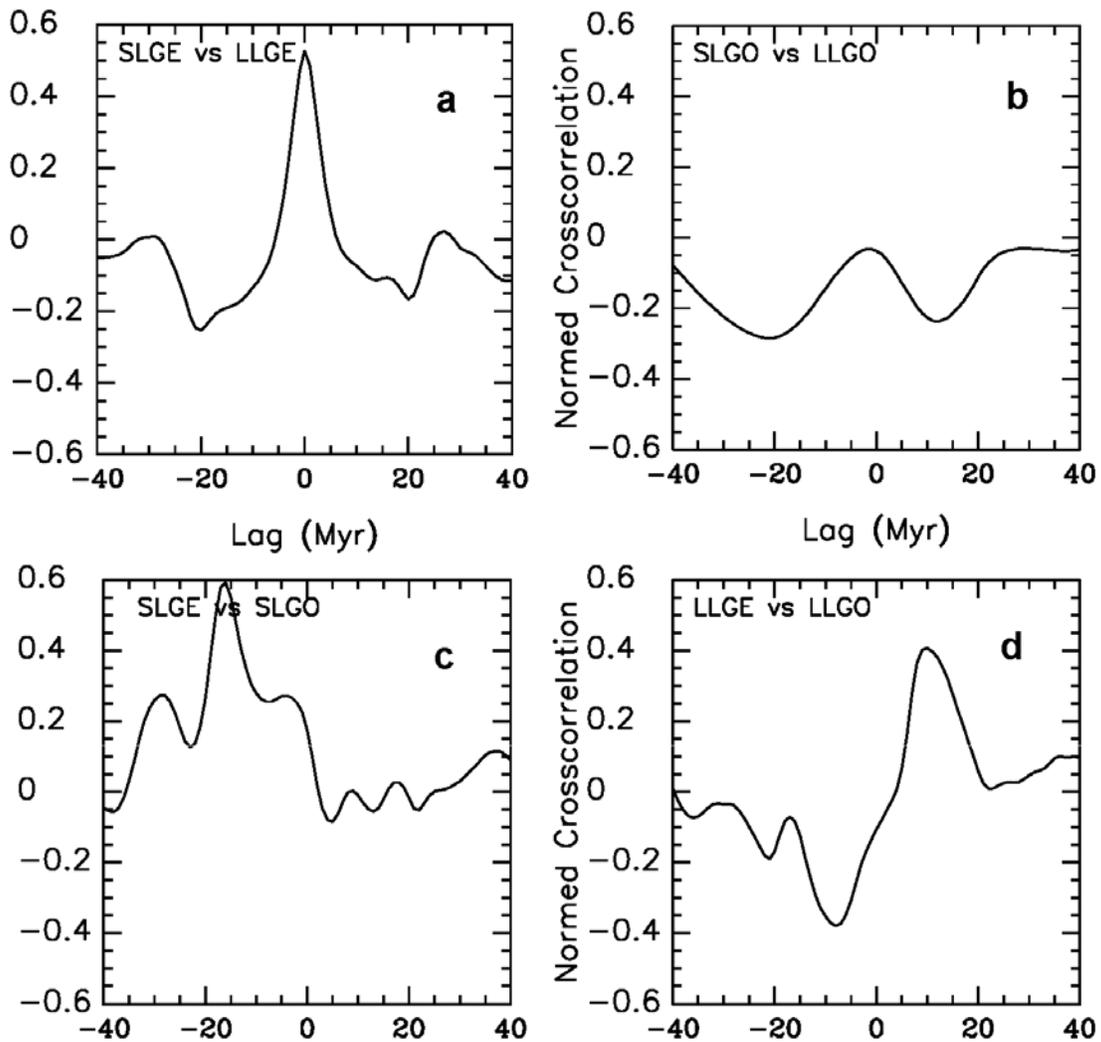

Figure 6: Cross-correlations of (A) short-lived genus extinction rate vs long-lived genus extinction rate and (B) short-lived genus origination rate vs long-lived genus origination rate. Panels (C and D) are normed cross-correlations of detrended extinction and origination rates, each divided by its standard deviation so that they contribute equally, for short-lived genera (C) and long-lived genera (D). The values given are based on series detrended and divided by their own standard deviation for normalization. The 95% confidence limits are very close to ± 0.2. The numbers reflect contributions from the entire spectrum, not just the 62 Myr fluctuation. In A extinction rates show a strong

peak at 0 lag, indicating that such events tend to affect them simultaneously. In B origination shows very little structure beyond a weak tendency for the two kinds of originations to avoid one another. In C short-lived genera show a strong correlation between their origination rates and extinction rates about 16 Myr later. Long-lived genera show in D a strong correlation between their extinction rates and origination rates about 10 Myr later, as well as an anticorrelation between their origination rates, and extinction rates about 8 Myr later.

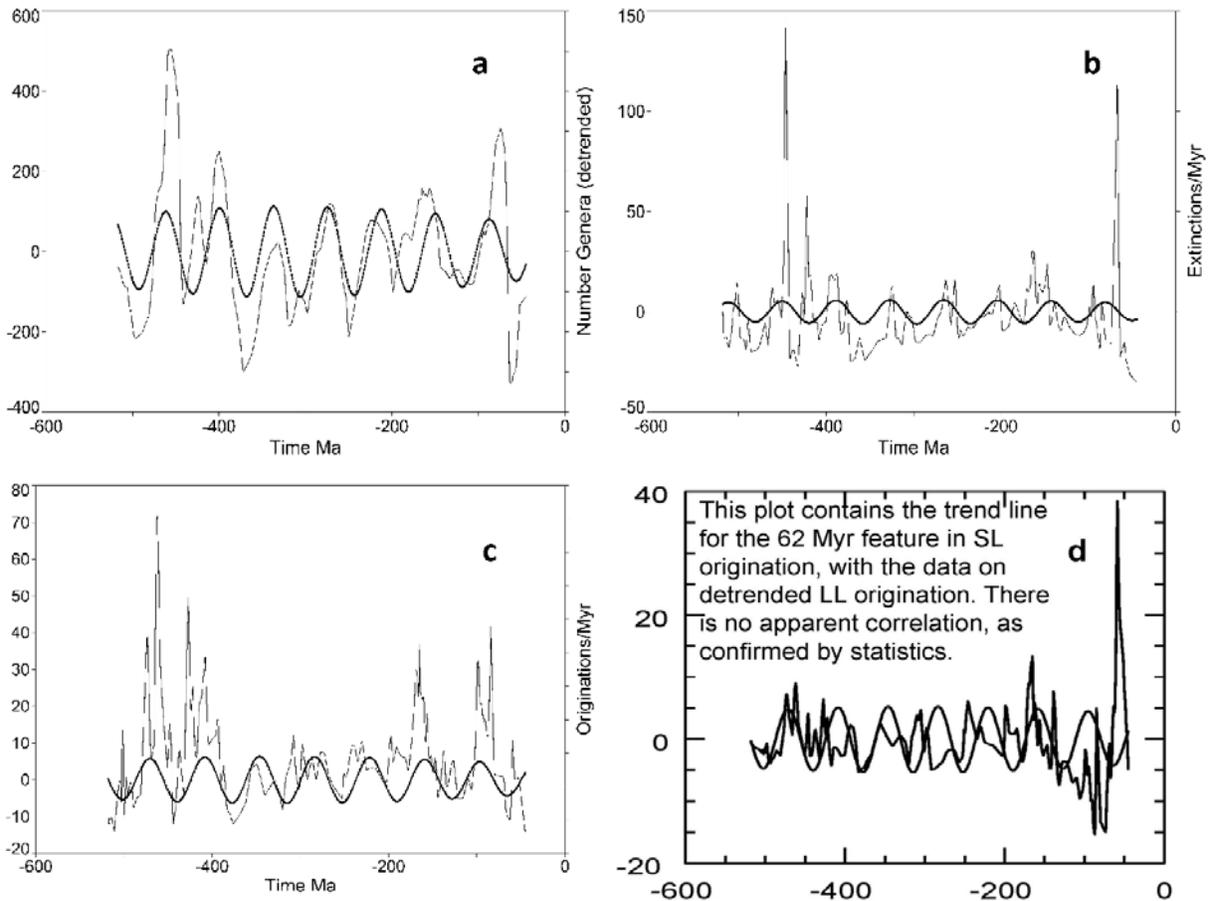

Figure 7:   Visual comparison of the fluctuations represented by spectral peaks with the underlying data. The amplitude of each sinusoid shows how much of the variance of the data is contained within the spectral peak. (A) Total short-lived genus diversity, corresponding to the peak in the spectrum displayed in Fig. 1B. The amplitude of the wave is large, and the overlap with the parent curve is readily apparent. (B) SLG extinction rate, corresponding to the spectrum shown in Fig. 3B. The sinusoid has lower amplitude but upon inspection has a correlation with the parent curve. (C) SLG origination rate, corresponding to the spectrum shown in Fig. 3A. The level of correlation is again apparent. (D) In order to investigate any relationship between long-lived genus origination and the ~62 Myr periodicity, we plot the long-lived genus

origination rate against the best-fit sinusoid for the short-lived genera. There is no obvious correlation.

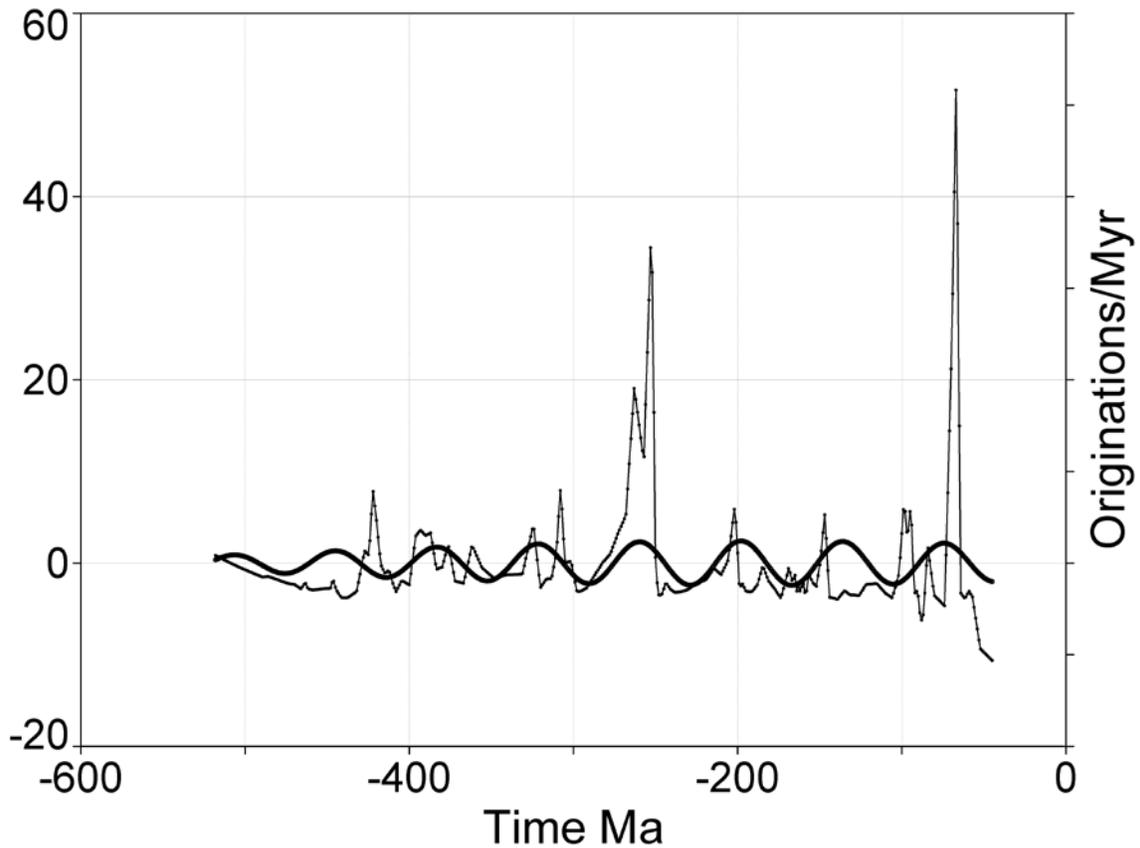

Figure 8:  The weak feature at 62 Myr in the origination rate for long-lived genera is used to generate a sinusoid plotted against the parent data. Some correlation is present. An approximate 20 Myr displacement of this sinusoid with respect to that shown in Figure 7C is consistent with the general sense of the result in Figure 6B.

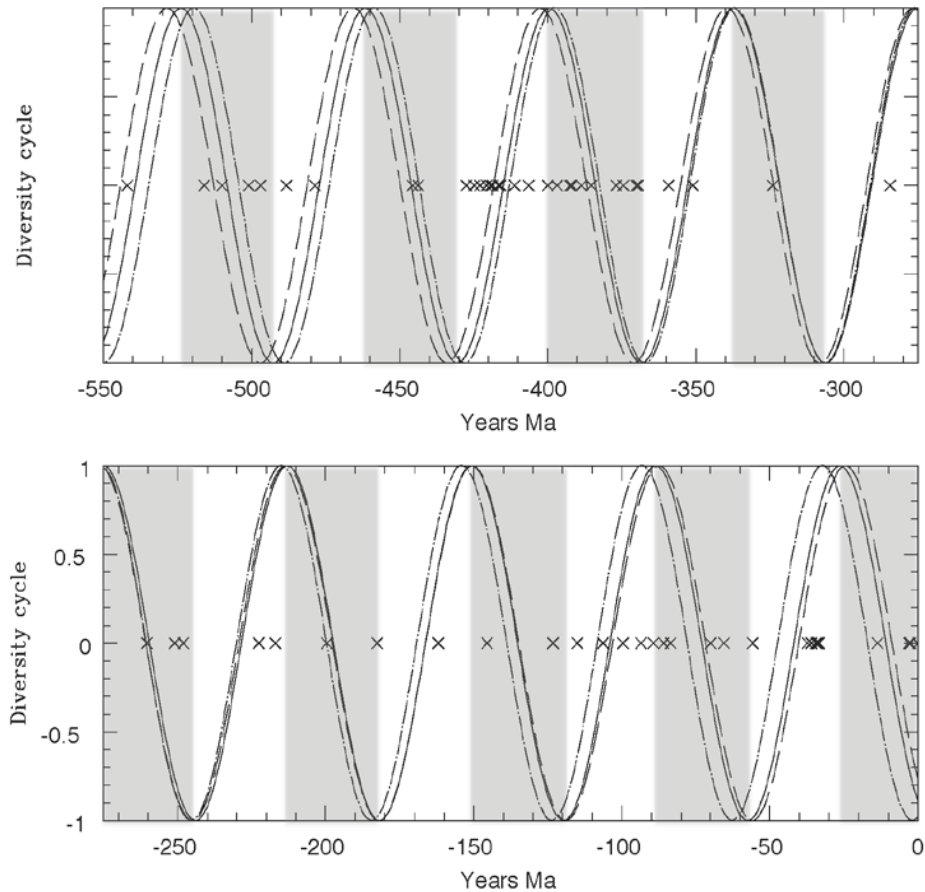

Figure 9A. The fits correspond to the 63 Myr component of the PBDB data (dashed line), the 62 Myr component of the R&M fit to the Sepkoski data (solid line), and the 61 Myr component of the FR2 marine genera (dash-dot line). Gray regions correspond to times of declining fossil biodiversity in the 62 Myr component of the R&M fit. The x's mark the 62 extinction events identified in summary listings noted in the text. They are nearly equally divided between times of declining and increasing biodiversity.

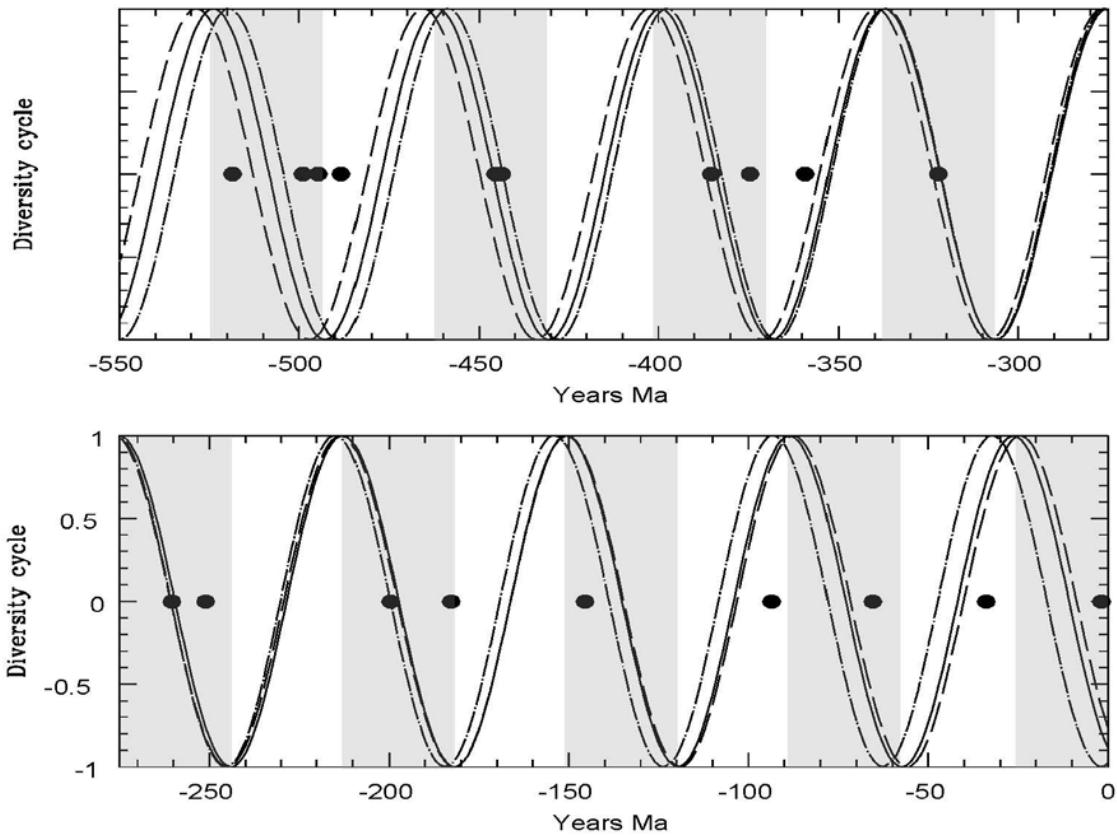

Figure 9B: The fits correspond to the 63 Myr component of the PBDB data (dashed line), the 62 Myr component of the R&M fit to the Sepkoski data (solid line), and the 61 Myr component of the FR2 marine genera (dash-dot line). Gray regions correspond to times of declining fossil biodiversity in the 62 Myr component of the R&M fit. The filled black circles are the times of extinction events identified by Bambach (2006). These events occur significantly more often than random during times predicted to be declining diversity by the ~62 Myr spectral component of any of the three data sets.

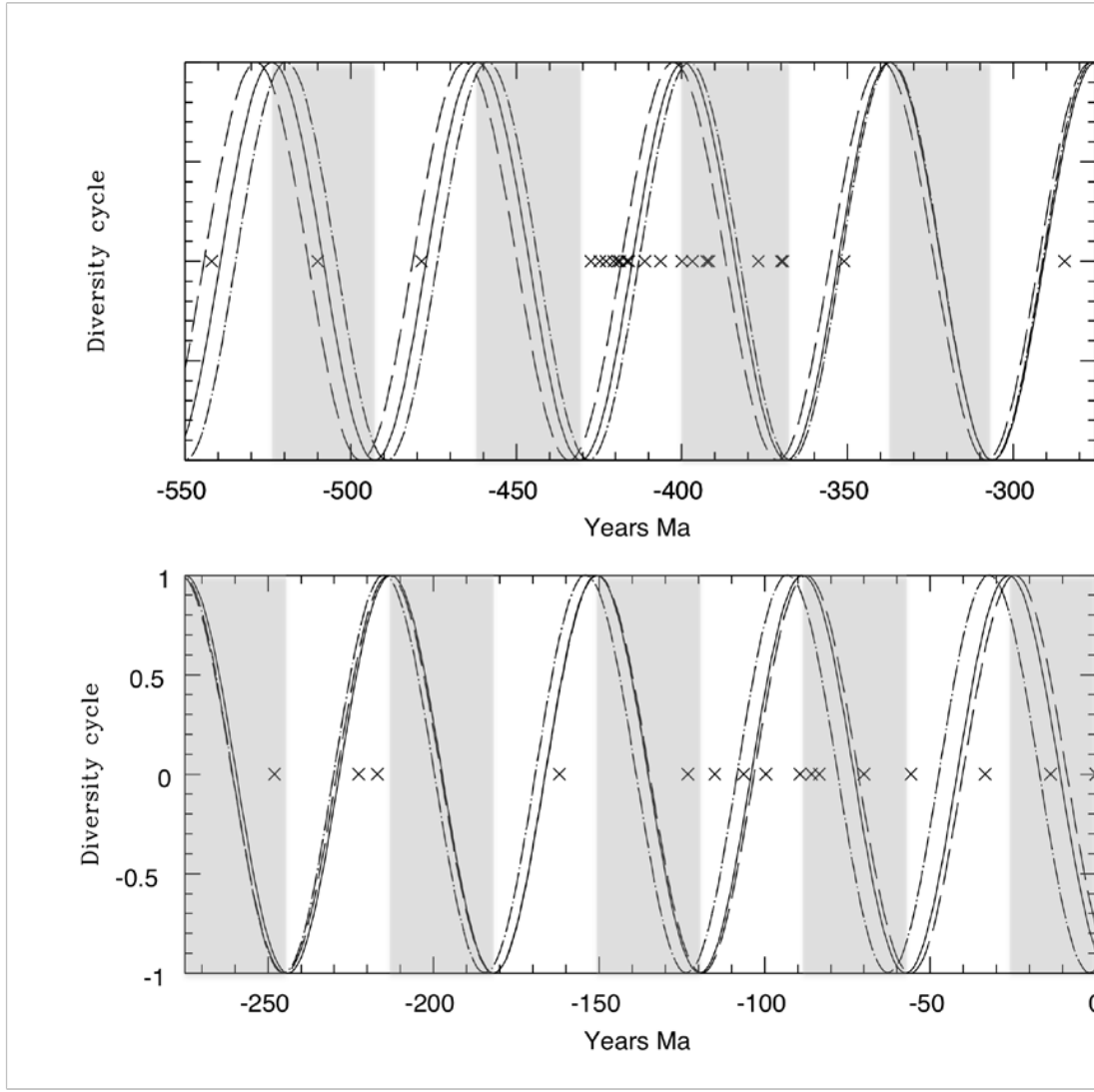

Figure 9C. The fits correspond to the 63 Myr component of the PBDB data (dashed line), the 62 Myr component of the R&M fit to the Sepkoski data (solid line), and the 61 Myr component of the FR2 marine genera (dash-dot line). Gray regions correspond to times of declining fossil biodiversity in the 62 Myr component of the R&M fit. The x's are extinction events that do not qualify as mass extinctions using the criteria of Bambach (2006), which with marginal significance prefer times of *increasing* biodiversity.

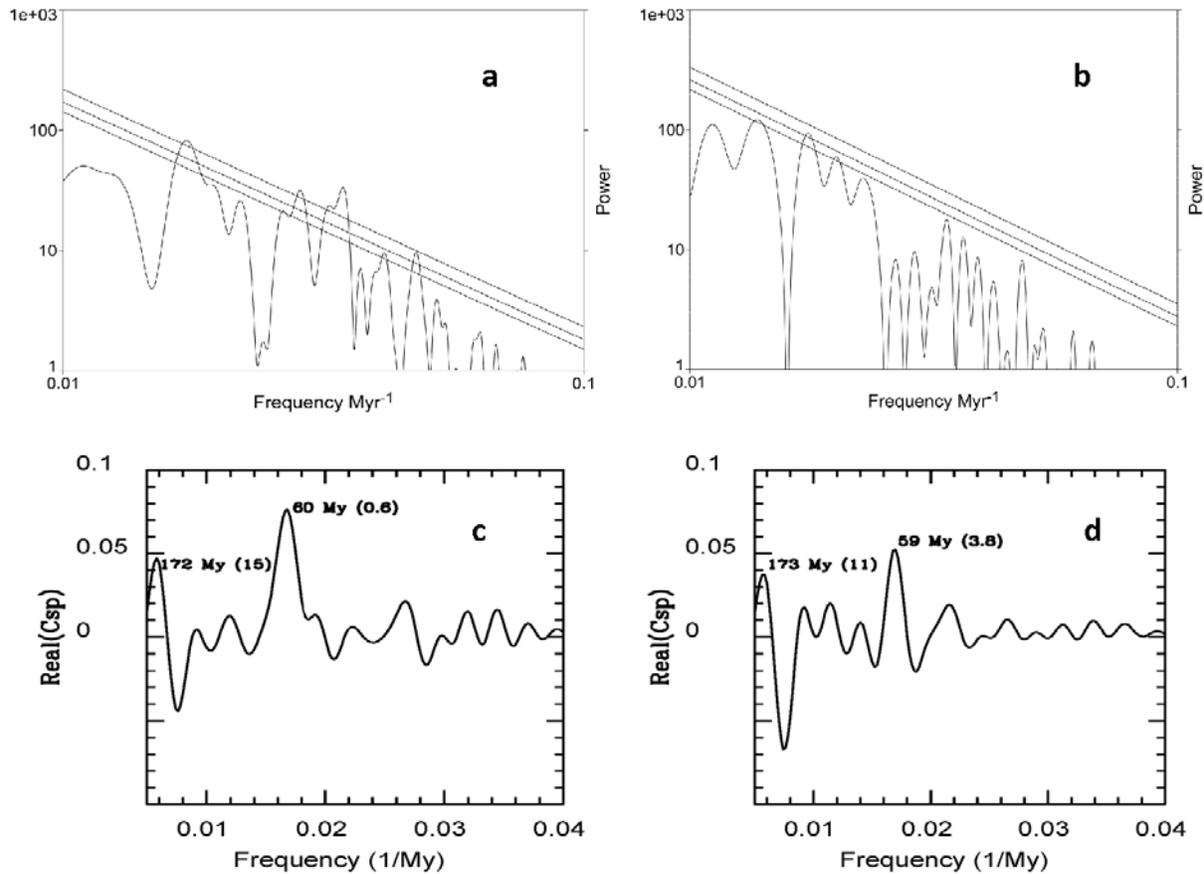

Figure 10:  (A) The power spectrum of the rate of deposition of carbonate-based rock packages as a function of time, based on data from Peters (2008). The parallel lines indicate significance at levels p = 0.05, 0.01, and 0.001 against the probability of any such peak arising against the spectral background. There is a spectral peak at p<0.001 at 58 ± 4 My, which accounts for 22% of the variance of the detrended data. The higher frequency peaks cannot be trusted unless independently confirmed, because of the relatively poorer time resolution of this data set with 63 points. (B) The power spectrum of siliciclastic packages, also based on Peters (2008) data.  There is a weak peak (p = 0.01) at 58 ± 3 Myr which accounts for 10% of the variance of the detrended data.

(C) The real part of the cross-spectrum between the detrended R&M fossil biodiversity and carbonate rock packages as described in Peters (2008).  The same two peaks seen before appear, but with slightly lower amplitude.  The degree of timing agreement between the two datasets is high on a 60 Myr peak, less so on a lower 172 Myr peak.  Since the area under these curves and above zero is proportional to the cross-correlation in the two series, we can see that most of it in fact comes from these spectral components, suggesting a common relationship at this frequency.  However, there is considerable negative area under the curve at a period of 135 Myr, so the overall crosscorrelation will be weak. These longer periods may be sensitive to detrending procedures (see Paper I) and require further study. (D) Another cross-spectrum, now between the Rohde and Muller (2005) time series and Peters (2008) siliciclastic rock packages.  The peaks appear, but now with lower amplitude and lessened timing agreement with fossil biodiversity than carbonate packages.  Again there is a positive peak near the frequency of interest, but given another, stronger, negative peak at 135 Myr the integrated area will be closer to zero than it was with the carbonate packages.

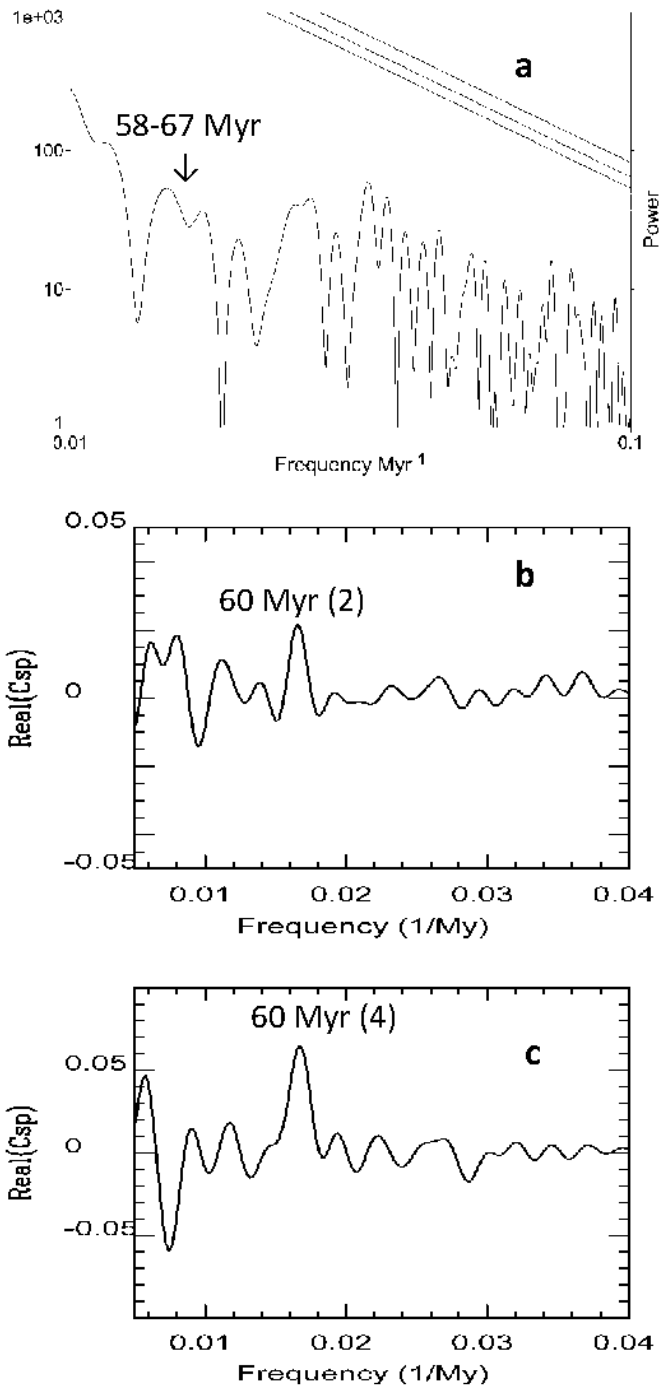

Figure 11: (A) The power spectrum of fluctuations in the detrended Timescale Creator sea level data. There are no significant spectral peaks indicating significant particular periodicity with this range of 25 to 200 Myr cycles. The nearest peak to the 62 Myr peak

of interest is a very broad feature across 58-67 Myr, so the frequency agreement is not particularly good. It accounts for less than 1% of the variance of this detrended data.  Its phase angle is, however, reasonably coincident with a component having a connection with biodiversity.  The cross-spectra will provide additional information. (B) The real part of the cross-spectrum between the TSC sea level data as above and the R&M biodiversity data.  Note that the zero-point of the y axis is shifted from cross-spectra previously shown.   There is a weak positive fluctuation at 60 Myr. This indicates a weak agreement between this sea level and biodiversity fluctuation component, since the phase agreement is very good. (C) Real cross-spectrum between the TSC sea level data and the sum of siliciclastic and carbonate rock packages (as noted in the text, the combination better represents the full suite of sediments that might be affected by sea level).  There is a weak peak at 60 Myr, again with good phase agreement.  This is less than one standard deviation different from the 62 Myr frequency of the biodiversity data peak value, in their joint fluctuation.

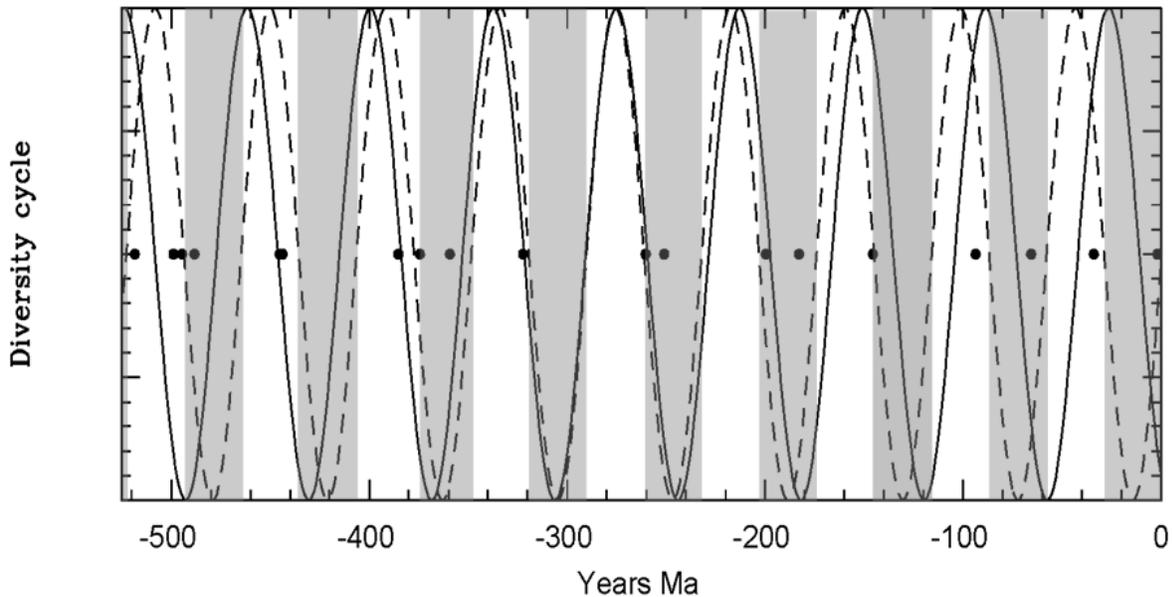

Figure 12*:* The fit to the 63 Myr component of the original R&M study (solid line) is plotted together with the fit to the 58 Myr component of the sedimentary rock data (dashed line).  The curves are somewhat out of phase near the present, but come into phase in the past, reaching peak agreement near the time of the end-Permian extinction.  The mass extinction events (Bambach 2006) do not preferentially occur when rock deposition rates are low (shaded areas), but rather tend to happen (p=0.03) when they are declining (even if currently high).  This argues for some common causal mechanism, rather than a selection bias in the assignment of extinction dates based on available sediment.

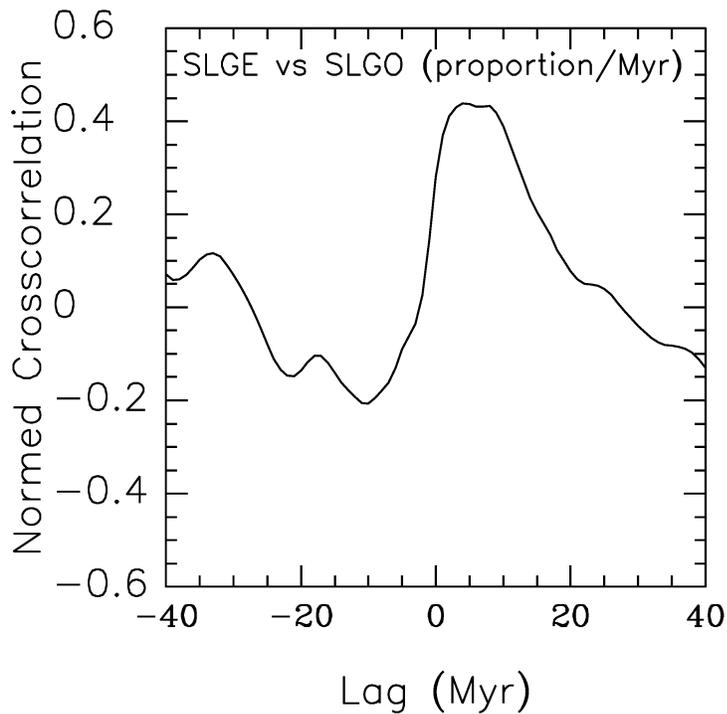

Figure 13: The cross-correlation of the proportion of extinction of short-lived genera per Myr against the proportion of origination. The 95% confidence intervals are close to ± 0.2. This shape is completely different from Figure 6c, which is based on the rates. Proportional normalization shifts the results because prior extinction reduces the number, increasing the proportion and the result at positive lag, and prior origination increases the number, reducing the proportion and the result at negative lag.